%% file: main.tex
\documentclass[twocolumn]{aastex63}

\newcommand\notready[1]{}

\usepackage{color}
\newcommand\editremark[1]{{\color{red}#1}}
\newcommand\ZD{}

\newcommand\E[1]{{\left\langle #1 \right \rangle}}

\newcommand\rate{\mathcal{R}}

\newcommand\param{\lambda}
\newcommand\Param{\Lambda}

\usepackage{amsmath,amssymb,mathtools}
\usepackage{graphicx}
\usepackage{lineno}

\graphicspath{{./}{figures/}}

\accepted{March 11, 2020}
\published{April 1, 2020}
\submitjournal{ApJ}

\shorttitle{Hierarchical Mergers}
\shortauthors{Doctor, Wysocki, et al.}

\begin{document}

\title{Black Hole Coagulation: Modeling Hierarchical Mergers in Black Hole Populations}
\author{Z.~Doctor}
\altaffiliation{zoheyr.doctor@gmail.com}
\altaffiliation{NSF GRFP Fellow}
\affiliation{Department of Physics,
University of Oregon,
Eugene, OR 97403, USA}
\affiliation{Department of Physics,
University of Chicago,
Chicago, IL 60637, USA}
\affiliation{Kavli Institute for Cosmological Physics,
University of Chicago,
Chicago, IL 60637, USA}
\affiliation{Enrico Fermi Institute,
University of Chicago,
Chicago, IL 60637, USA}

\author{D.~Wysocki}
\affiliation{Center for Computational Relativity and Gravitation, Rochester Institute of Technology, Rochester, New York 14623, USA}

\author{R.~O'Shaughnessy}
\affiliation{Center for Computational Relativity and Gravitation, Rochester Institute of Technology, Rochester, New York 14623, USA}

\author{D.~E.~Holz}
\affiliation{Department of Physics,
University of Chicago,
Chicago, IL 60637, USA}
\affiliation{Department of Astronomy and Astrophysics,
University of Chicago,
Chicago, IL 60637, USA}
\affiliation{Kavli Institute for Cosmological Physics,
University of Chicago,
Chicago, IL 60637, USA}
\affiliation{Enrico Fermi Institute,
University of Chicago,
Chicago, IL 60637, USA}

\author{B.~Farr}
\affiliation{Department of Physics,
University of Oregon,
Eugene, OR 97403, USA}

\begin{abstract}
Data from the LIGO and Virgo detectors has confirmed that stellar-mass black holes can merge within
a Hubble time, leaving behind massive remnant black holes.  In some
astrophysical environments such as globular clusters and AGN disks, it may be possible for these remnants
to take part in further compact-object mergers, producing a population of hierarchically formed black holes.  In
this work, we present a parameterized framework for describing the population of binary black hole mergers, while
self-consistently accounting for hierarchical mergers.  The framework casts black holes as particles in a box which can collide
based on an effective cross-section, but allows inputs from more detailed astrophysical simulations.  Our approach is
relevant to any population which is comprised of second or higher generation black holes, such as primordial
black holes or dense cluster environments.
We describe some possible inputs to this generic model and their effects on the black hole
merger populations, and use the model to perform Bayesian inference on the catalog of black holes from
LIGO and Virgo's first two observing runs. We find that models with a high rate of hierarchical mergers are disfavored, consistent with previous population analyses.  Future gravitational-wave
events will further constrain the inputs to this generic hierarchical merger model, enabling a deeper look
into the formation environments of binary black holes.
\end{abstract}

\keywords{editorials, notices ---
miscellaneous --- catalogs --- surveys}





\section{Introduction}

\input{intro}

\section{Parameterized hierarchical formation of binary black holes}
\label{sec:model}

\input{model}

\section{Constraining hierarchical formation with gravitational wave observations}
\label{sec:infer}

\input{inference}

\section{Discussion}
\label{sec:discuss}

\input{discussion}

\section{Conclusions}
\label{sec:conclude}

\input{conclusion}

\acknowledgments
ZD, ROS, and DEH initiated this work at the Kavli Institute for Theoretical Physics, supported in part by NSF PHY-1748958. ZD was supported by the NSF Graduate Research Fellowship Program, grant DGE-1144082. ROS and DW are supported by NSF AST-1707965 and PHY-1909534. ZD and DEH are supported by NSF grant PHY-1708081 and the Kavli Institute for Cosmological Physics at the University of Chicago through an endowment from the Kavli Foundation.  DEH also gratefully acknowledges support from the Marion and Stuart Rice Award. BF is supported by NSF grant PHY-1807046.
The authors thank the
LIGO Scientific Collaboration for access to computing resources and
public data and
gratefully acknowledge the support of the United States
National Science Foundation (NSF) for the construction
and operation of the LIGO Laboratory and Advanced
LIGO as well as the Science and Technology Facilities
Council (STFC) of the United Kingdom, and the Max Planck-Society (MPS) for support of the construction of
Advanced LIGO. Additional support for Advanced LIGO
was provided by the Australian Research Council.
For their use in implementing our hierarchical inference code, we would like to acknowledge \textsc{Numpy} and \textsc{Scipy} \cite{scipy}, \textsc{Matplotlib} \cite{matplotlib}, and \textsc{h5py} \cite{collete_2013}.

\appendix

\section{Semianalytic approach to hierarchical mergers}
\input{appendix_semianalytic}
\section{Changing the merging fraction}
\label{app:merge_frac}
\input{merge_frac.tex}

\bibliography{references.bib,LIGO-publications}

\end{document}

%% file: intro.tex

The Advanced Laser Interferometer Gravitational Wave Observatory (LIGO) ~\citep{2015CQGra..32g4001T} and  Virgo
\citep{gw-detectors-Virgo-original-preferred} 
detectors have
and will continue to discover gravitational waves (GW) from coalescing binary black holes (BBHs) and neutron stars.  So
far, several
tens of binary black hole detection candidates have been reported in O3, LIGO's current observing run, and several hundreds more detections are expected over the next five years  \citep{RatesPaper,LIGO-O1-BBH}.
As the cosmic census these surveys provide grows more comprehensive, these observations will discriminate between
formation scenarios of compact-object binaries.
\citep{2010CQGra..27k4007M,gwastro-popsynVclusters-Rodriguez2016,2016ApJ...830L..18B,2016PhRvD..94f4020N}. 
A few formation scenarios invoke ``hierarchical'' growth of binary black holes in which some black holes
are themselves products of previous  mergers.
These hierarchical mergers could occur in
globular clusters \citep{2002ApJ...576..899P,2006ApJ...640..156G}, 
AGN disks  \citep[see,e.g.][]{2012MNRAS.425..460M,2017ApJ...835..165B,2019arXiv190609281Y,gwastro-agndisk-McKernanPredictMassSpin2019}, or
nuclear star clusters \citep{2016ApJ...831..187A}. Alternatively, the hierarchical merger components could have been produced in the early
universe due to primordial density fluctuations forming primordial black holes \citep{2015PhRvD..92b3524C,2017PDU....15..142C,Belotsky}.
Notably, hierarchical growth produces distinctive signatures in the mass and spin distribution \citep{2017ApJ...840L..24F,2019arXiv190609281Y,GerosaBerti,gwastro-agndisk-McKernanPredictMassSpin2019,2019arXiv190307813K}, the most generic of which is a population of spinning black holes.   
For some realizations of these models' parameters, several groups have made predictions about the black hole mass and
spin distribution  \citep{gwastro-popsynVclusters-Rodriguez2016,gwastro-agndisk-McKernanPredictMassSpin2019,popsyn-KB-LowNatalBHSpin-2017}.
Additional investigations have assessed whether existing observations are compatible with these models, focusing on the individual
event GW170729 \citep{gwastro-170729HM-Katerina,2019arXiv190609281Y,2019arXiv190307813K}.

In this work, we introduce a generic, parameterized framework that accounts for binary black holes which form through hierarchical
mergers.  The method treats black holes as particles in a box which undergo collisions based on an effective cross section.
This framework can incorporate a wide range of submodels and prescriptions, enabling one to create models that are purely phenomenological
or instead heavily based on detailed astrophysical investigations and simulations.    
We provide a concrete implementation of our framework, including astrophysically realistic initial conditions.
Using existing gravitational wave observations, we perform Bayesian inference on our parameterized model.
%


Our paper is organized as follows.
In \S\ref{sec:model}, we describe our framework for hierarchical mergers and some parameterizations within the framework, illustrating them with
simple examples.  We also describe our fiducial initial conditions for binary black hole populations.  
In \S\ref{sec:infer}, we show how to constrain this parameterized model through comparison with gravitational wave observations from LIGO and Virgo's
first and second observing runs.
In \S\ref{sec:discuss}, we discuss the results of our parameter inference on the LIGO-Virgo data, the overall efficacy of our framework, and possible
extensions to the parameterizations explored herein.
Finally, we summarize the results of our investigation in \S\ref{sec:conclude}.

%% file: model.tex

\subsection{General framework}

We employ a flexible method for self-consistently generating  mass and spin distributions for binary black holes which
include a subpopulation of hierarchical mergers.   Rather than model the complex dynamics of individual stellar
environments, we build a parameterized phenomenological
model which describes the aggregate properties of merging binaries in the local universe, using volume-averaged coupling
coefficients.  Our framework incorporates three generic  physical processes.
First, black holes {\it coagulate} when pairs of compact objects merge into single compact
  objects which may remain in the population.
Second, we  allow for {\it depletion}, where some compact objects leave dense environments and no longer have
  an opportunity to merge with other objects.
Finally, we allow for {\it augmentation}, where some process introduces new
compact objects to the hierarchical interacting environment (e.g., BHs from stellar collapse or AGN disk dynamics).

\begin{widetext}
Following similar investigations \citep{Christian18,Lissauer1993}, we model these effects 
with a Monte Carlo procedure, designed to approximate a continuous-time coagulation equation \citep{Smoluchowski1916}, which has the qualitative form
\begin{align}\label{eq:coag}
\partial_t f(x;t) &= \frac{1}{2} \int dx'dx'' f(x';t)f(x'';t) \Gamma(x',x'';t)\delta(x_{\mathrm{rem}}(x'',x') - x) 
-   \int dx'  f(x;t) f(x';t) \Gamma(x,x';t) \\
&+r(x;t) - d(x;t) \nonumber
\end{align}
where here $x$ denotes black hole parameters, $f(x;t)$ denotes the BH parameter distribution function at time $t$, $\Gamma(x,x';t)$
denotes a volume-averaged interaction rate (i.e.~coagulation), and $r(x;t)$ and $d(x;t)$ are the augmentation and depletion rates of black holes with
parameters $x$ at time $t$. The first integral describes the accumulation of black holes with parameters $x$ due to mergers of pairs
of black holes with parameters $x',x''$.  The delta function enforces that the final parameters $x$ are produced by a merger of BHs
with parameters $x',x''$. The function $x_{\rm rem}(x,x')$ computes the remnant parameters from merger component parameters $x$ and $x'$
\footnote{If the remnant mass of merging black holes was exactly the sum of the merging components, then $m_{\rm rem}(m,m')=m+m'$, but since energy
is radiated in gravitational waves from the coalescence, $m_{\rm rem}(m,m') < m+m'$.}.  
The second integral accounts for the decrease of black holes with parameters $x$ due to mergers with other black holes with parameters $x'$, 
and its integrand $f(x;t) f(x';t)\Gamma(x,x';t)$ is equivalent to the merger rate as a function of parameters.   
In the absence of augmentation or depletion, the total number of black holes $\int f dx$ decreases as $-1/2 \int dx dx'
\Gamma(x,x';t) f(x;t) f(x';t)$, as each merger reduces the total number of black holes by one.  (The factor of $1/2$ is a
statistical factor to avoid overcounting.)
\end{widetext}

Given an initial condition $f(x,t_0)$, an interaction rate $\Gamma(x,x';t)$, a map between merger components and remnants $x_{\rm rem}(x,x')$, and prescriptions
for augmentation and depletion, the solution $f(x;t)$ can in principle be computed.  This approach is highly modular and can 
incorporate  complex dynamical physics via the coagulation, augmentation, and depletion functions. Additionally, existing black hole population
models can be extended to include hierarchical merger effects in our framework.  With this framework in hand, we 
first describe our method for computing these hierarchical merger distributions and then
turn to astrophysically motivated choices for these functions and their application to GW data.

\subsection{Monte Carlo Implementation}\label{subsec:MC}
To solve Equation \ref{eq:coag}, we perform an iterative procedure on a sample of black holes. First, a ``natal'' black hole sample is chosen, i.e.~samples
from $f(x,t_0)$.
Then at each step, a set fraction $w$ of the black holes
are merged based on the coagulation coupling, and the final mass, spin, and kick velocity are computed for the merger remnants.  
The kick velocities of 
these remnant black holes determine whether they are reintroduced to the overall sample of black holes or if they are removed due to leaving the environment.  
Meanwhile, new black holes formed from non-hierarchical processes can be added to the sample. The fraction that are merged at each iteration is a proxy
for the timescale on which these mergers can occur.  If the fraction is small, few mergers will occur at each iteration, but
the mergers that do occur will have the opportunity to merge again in the next iteration, allowing more unequal-generation mergers.
This approximates continuous coagulation.
If on the other hand the fraction is order unity, most of the black holes will merge during each time step.  In the
latter scenario, the
black holes in the sample will typically be of the same generation at each time step, as if some process delayed their re-entrance to the population
immediately after coagulation.  Here we fix this fraction $w$ to 5\%, as a large timestep which still reasonably
approximates continuous evolution;  we expand on the fraction size in
Appendix \ref{app:merge_frac} and note that future work could allow this 
to be a free parameter.  We summarize our full Monte Carlo procedure below:
\begin{enumerate}
  \item Sample $N$ black holes from the natal population.  Each BH has a mass and spin parameter.  Call this sample
    $S$.
  \item Pair $wN$ black holes randomly from $S$, weighted by the coagulation coupling prescription, where $w$ is the fraction of BHs that
    merge at each iteration.
  \item Compute the final mass, spin, and kick velocity for the black hole pairs to create a new sample of post-merger black holes 
    called $S'$ and remove any black holes that were paired from $S$.  
  \item Remove black holes from $S'$ based on their kick velocities using a model for black hole depletion.
  \item Sample more black holes based on the augmentation prescription and call this sample $S''$.
  \item Set $S = S \cup S' \cup S''$.
  \item Repeat steps 2-6 until the maximum number of desired iterations is reached. 
\end{enumerate}


\subsection{Model Prescriptions and Parameterizations}
In this section, we describe our inputs to Equation \ref{eq:coag}, which we have chosen to be simple, computationally efficient, 
and astrophysically motivated. 
\ZD{Notably, the choices we make here all assume an isotropic interaction environment, with randomly oriented spins, which
which may not be well-suited to some environments such as AGN disks.}
However, we emphasize that alternative effects can be readily incorporated into this framework if desired.  
To limit the scope of our investigations, augmentation is not considered in this work,
but future studies could include it.

\subsubsection{Coagulation}
For simplicity, we assume the volume- and time-averaged interaction rate $\Gamma$ depends only on binary masses $(m,m')$, with a parametric form
\begin{eqnarray}
\label{eq:gamma}
\Gamma_{m,m'} \propto \left(\frac{(m+m')}{M_{\rm ref}}\right)^a \left(\frac{\eta}{\eta_{\mathrm{ref}}}\right)^b
\end{eqnarray}  
(This single interaction term is designed to capture the average effect of interactions throughout the volume, on
the long timescales over which the BH mass distribution evolves appreciably through hierarchical mergers.)
We include the total binary mass dependence $\left(\frac{(m+m')}{M_{\rm ref}}\right)^a$  for two reasons. Firstly,
 bigger black holes have larger ``cross-sectional areas'' with which they can interact with other objects.  In the limiting
case of spheres in a gas with radii $r$, one would expect $\Gamma(r,r') \propto (r+r')^2$.  To account for the complex dynamics
of interacting black holes, we do not fix the power to 2 and instead let it vary, and since a black hole's Schwarzchild radius is
 directly proportional to its mass, we replace radii with masses. The second effect this term accounts for is dynamical friction,
 which brings more massive black holes to dense centers of clusters where they can merge.  
 The second term $\left(\frac{\eta}{\eta_{\mathrm{ref}}}\right)^b$ depends 
 on the symmetric mass ratio $\eta$ to account for a possible preference for mergers to choose more equal or unequal masses \citep[see e.g.~][]{PickyPartners}.
 In globular clusters for example, mass segregation may favor equal-mass mergers over unequal mass
\citep{2008MNRAS.388..307S,2017MNRAS.469.4665P,2019PhRvD.100d3027R}. 

Although we assume that the black hole spins do not influence the interaction rate, we do keep track of 
the spin magnitudes of the black holes and calculate final black hole spins from initial component parameters.  We use 
fits to numerical relativity simulations from \citet{TichyMarronetti} for $x_{\rm rem}(x,x')$, the final mass and spin of a 
remnant black hole given the masses and spins of the individual components.   To further simplify our calculations, we assume 
the hierarchical environment is isotropic, so only spin magnitudes $\chi$ need to be tracked since spin orientations are random.
As such, we can simply write $x=(m,\chi)$ in this prescription.

\subsubsection{Depletion}\label{subsubsec:DepletionModel}
Remnant black holes experience recoil kicks which may eject the remnant from the environment and prevent it from 
merging again with another object.  Here we consider two cases: 1.~No depletion and 2.~cluster depletion. In the first
case, we assume no black holes leave the environment;  in the second, 
we use the ``$V_{459}$'' fits to numerical relativity simulations from \cite{ZlochowerLousto} for recoil velocities
with a prescription for the distribution of cluster escape velocities to calculate the depletion rate. 
\ZD{We parameterize the depletion based on the magnitude of the recoil velocity $v_{\rm kick}$, and ignore
the recoil direction, although future studies could incorporate the recoil direction to
account for anisotropy in the merger environment.} 
For cluster depletion, we assume that black holes are in star clusters with a variety of density profiles and hence
a variety of central escape velocities.  We write the escape probability as:
\begin{align}
\label{eq:pesc}
  p(&{\rm escape}| v_{\rm kick}, \mu_M,\sigma_M,\mu_{r_0},\sigma_{r_0}) \\
  &  \propto \int \int d\log{M} d\log{r_0} \Theta\left[\frac{1}{2}v_{\rm kick}^2 - \frac{GM}{r_0}\right]
\nonumber\\
&  \times\exp\left(-\left(\frac{\log M - \log \mu_M}{\sigma_M}\right)^2 - \left(\frac{\log r_0 - \log \mu_{r_0}}{\sigma_{r_0}}\right)^2\right)
\nonumber
\end{align}
The Heavyside function enforces that remnants with kick velocities larger than the cluster
escape velocity are ejected.  The cluster escape potential is given by a Plummer model
and the black holes are always assumed to be at the center of clusters. The last line of terms
describes the distribution of cluster masses $M$ and effective radii $r_0$ in the Plummer 
model.  We take these cluster masses and radii to be log-normally distributed \ZD{and parameterized
by $\mu_M$, $\sigma_M$, $\mu_{r_0}$, and $\sigma_{r_0}$}, but emphasize
that other choices could be made for all of these depletion prescriptions.

\subsubsection{Natal Populations}

\begin{figure}
\includegraphics[width=\columnwidth]{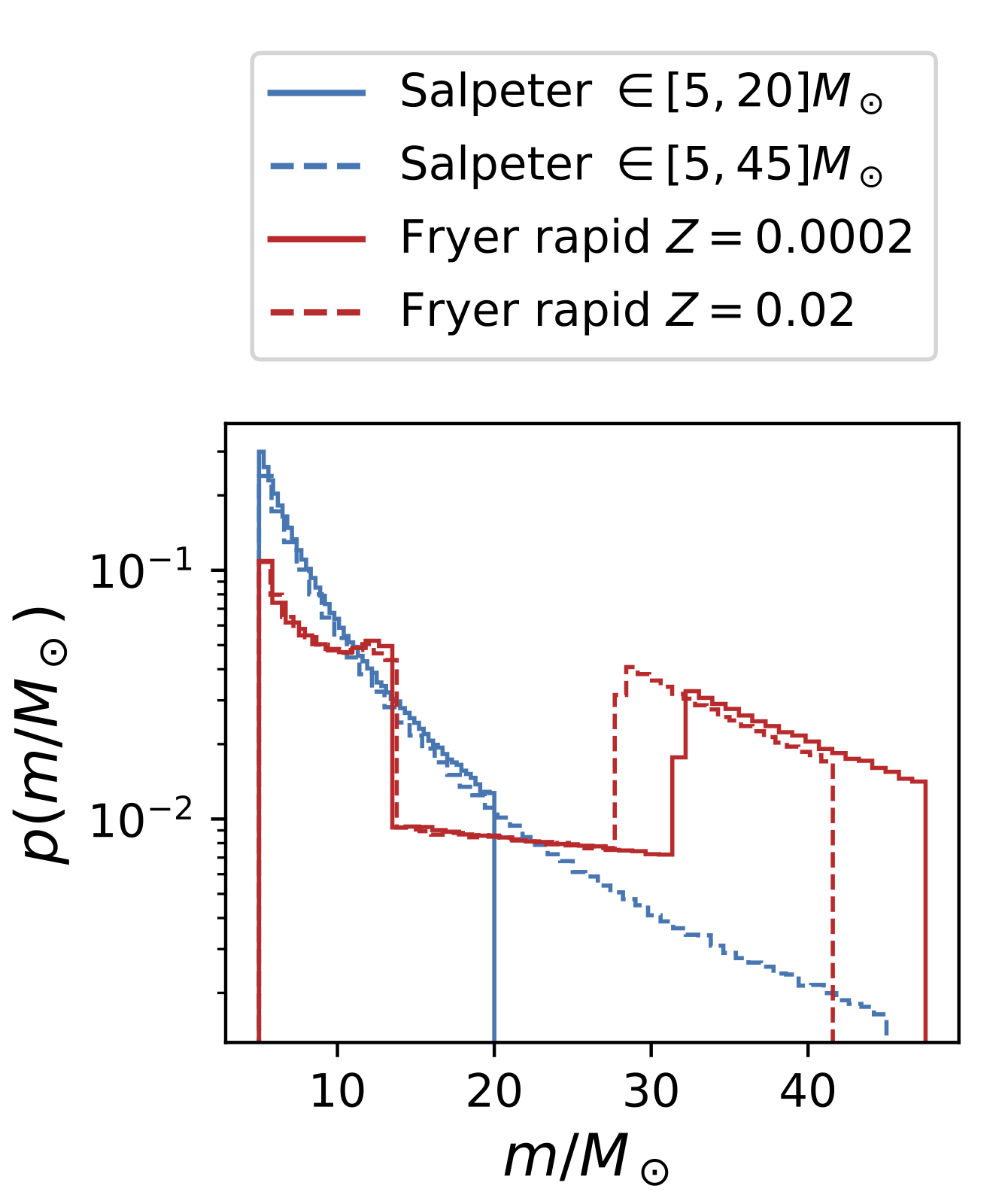}
\caption{
  Four different scenarios for initial black hole mass distributions.
  Blue curves denote a Salpeter-like powerlaw, with the solid (dashed) line
  corresponding to an upper mass cutoff of $20 M_\odot$ ($45 M_\odot$).
  Red curves denote the Fryer rapid model, with the solid (dashed) line
  corresponding to a metallicity $0.0002$ ($0.02$).
}
\label{fig:InitialConditionsAndImplications}
\end{figure}

The final ingredient we need to specify in our model is the initial distribution of masses and spins $f(x;t_0)$.  We
hereafter refer to this as the ``natal'' distribution, and take it to be the distribution of black hole parameters
formed at black hole birth.  A variety of choices could be made, but here we restrict ourselves to two cases. The first
case is a simple power-law mass function in component masses with lower and upper mass cutoffs
\begin{equation}
  p(m) \propto
  \begin{dcases*}
    m^{-\alpha}, & if $m_{\mathrm{min}} \leq m \leq m_{\mathrm{max}}$
    \\
    0, & otherwise.
  \end{dcases*}
  \label{eq:natal-powerlaw}
\end{equation}
In all cases we use the fiducial value $m_{\mathrm{min}} = 5 \, M_\odot$ for the sake of simplicity.  For the other
parameters, we either fix them to fiducial values of $\alpha = 2.35$ (from a Salpeter IMF) and $m_{\mathrm{max}} = 20 \,
M_\odot$ (from early stellar evolution modeling), or we allow the data to tune them, assuming uniform priors.  
The blue curves in Figure \ref{fig:InitialConditionsAndImplications} show two examples of black hole natal mass distributions with the Salpeter prescription.
The fiducial Salpeter mass distribution for black holes is a basic model which assumes that the fraction of mass retained from stellar birth to black hole formation, $m / m_{\mathrm{ZAMS}}$, is constant across all masses.  This is unlikely to be true in reality, as the processes undergone by a star depend strongly on its mass.  To take things a step further in complexity, we still assume the mass distribution follows a power law, but with an index $\alpha$ which differs from the IMF's value.  This is still fairly un-realistic, as the black hole natal mass spectrum is not expected to be this simple, but this at least lets the data determine the general trend of the spectrum.

Our second model has a better footing in physical principles, but loses some flexibility.  We assume a pure Salpeter IMF
for the ZAMS masses, in the range $[5, \infty) \, M_\odot$, and evolve them to black holes using the
  \cite{2012ApJ...749...91F} Rapid model.  (Our calculations implicitly adopt the same wind mass loss model as employed
    in that study.)   This introduces an additional hidden variable, the stellar metallicity
  $Z_{\mathrm{metal}}$ for each progenitor star.  The red and green curves in the bottom panel of    Figure
  \ref{fig:InitialConditionsAndImplications} show  our inferred progenitor distributions, for two choices of $Z$.  In principle, this should be a random variable, obeying some
  distribution which may correlate with the IMF.  For simplicity, however, and motivated by the approximate similarity
  between these two distributions,  we fix this to a constant
  $Z_{\mathrm{metal}}^*$, assumed to be the same for every progenitor.   

Now we turn to to the black hole natal spins.  Black hole natal spins remain a matter of considerable observational and theoretical debate.  Motivated by LIGO's
observations and recent modeling \citep{2019ApJ...881L...1F,popsyn-KB-LowNatalBHSpin-2017,FarrHolzFarr, O2pops}, we adopt a simple fiducial choice: all BHs in
our original population have small characteristic spin magnitudes, drawn from a Beta distribution with ${\rm mean}(\chi)=0.047$  and ${\rm Var}(\chi)=0.002$.
We also assume that the spin directions in the natal population are randomly oriented, but again we emphasize that other choices could be made.  

\subsubsection{Merger Rates}
As described in \S\ref{subsec:MC}, our Monte-Carlo procedure works with a finite set of black holes. We take these
black holes to be a proxy for the entire population and assume that the overall merger rate of black holes
is simply a scaled population of those generated in our Monte Carlo simulations.  We also stipulate that the 
merger rate density is constant in co-moving volume.  Future studies could certainly incorporate more detailed effects,
but here we opt for simplicity.  In the following section, we show normalized distributions of the masses and spins
of black holes, but in \S\ref{sec:infer} we present inference results that allow the merger rate density to be inferred
by the data.  



\subsection{Characterizing the parameters}
\label{sec:model:explain_params}
To elucidate the effect of each of the parameters described in the previous section,
we take the reader through a sequence of examples.  The examples we present here are
primarily for illustration and do not necessarily represent parameters that describe the 
observed population of black-hole mergers to date. Note that the histograms and kernel
density estimate curves shown here are not explicitly used in our analysis; they are simply
representations of the samples from our Monte-Carlo procedure.

\subsubsection{Time Evolution}

\begin{figure}
\includegraphics[width=\columnwidth]{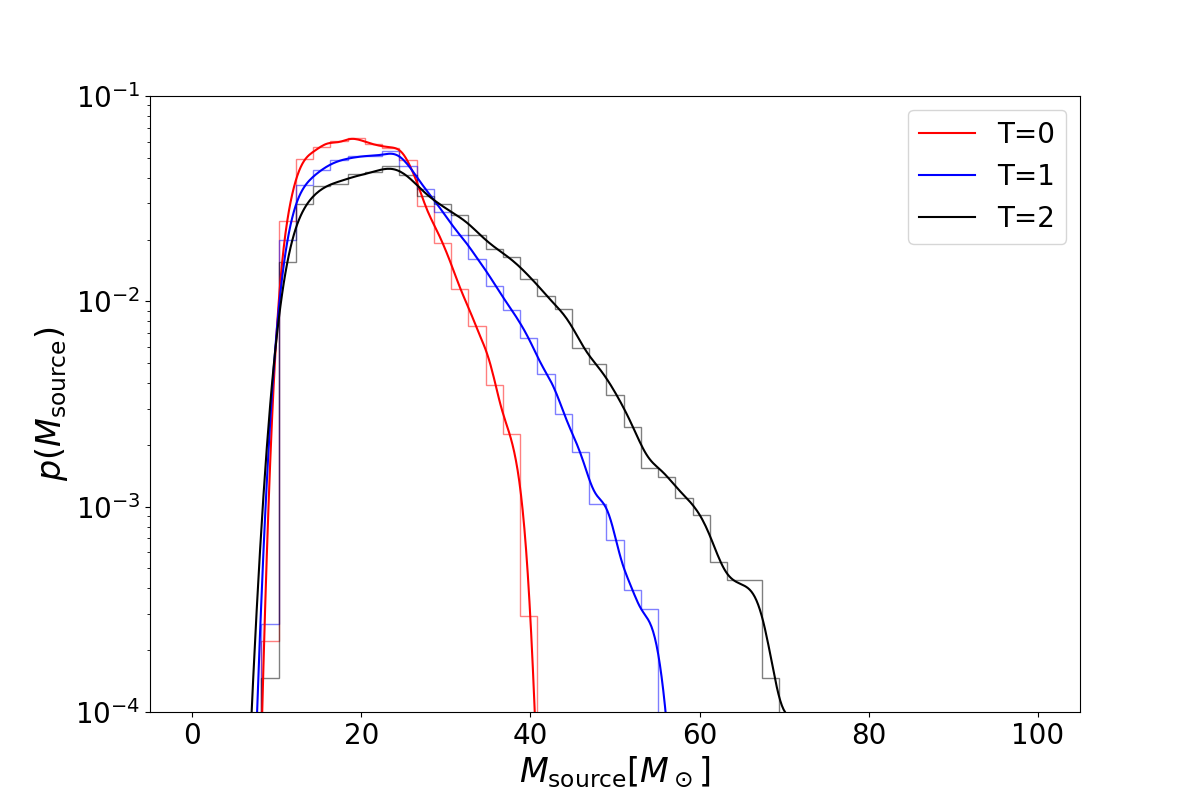}
\caption{\label{fig:timesteps} The total mass distribution of binary black hole mergers
at three successive time steps evolving from a Salpeter natal distribution ($\alpha=2.35$)
with coupling parameters $a=2$ and $b=0$.  The smooth curves overlaid on the
histograms are kernel density
estimates of the Monte-Carlo samples and are shown purely to guide the eye.}
\end{figure}

\begin{figure}
\includegraphics[width=\columnwidth]{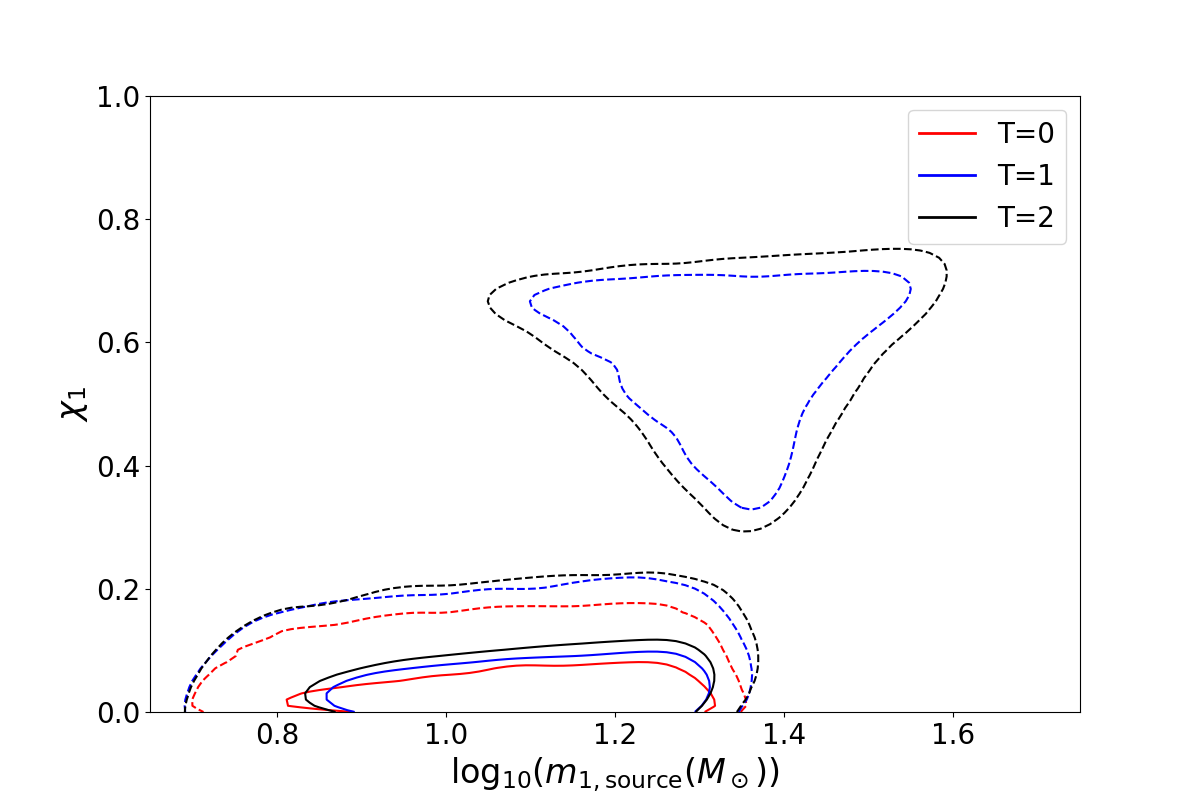}
\includegraphics[width=\columnwidth]{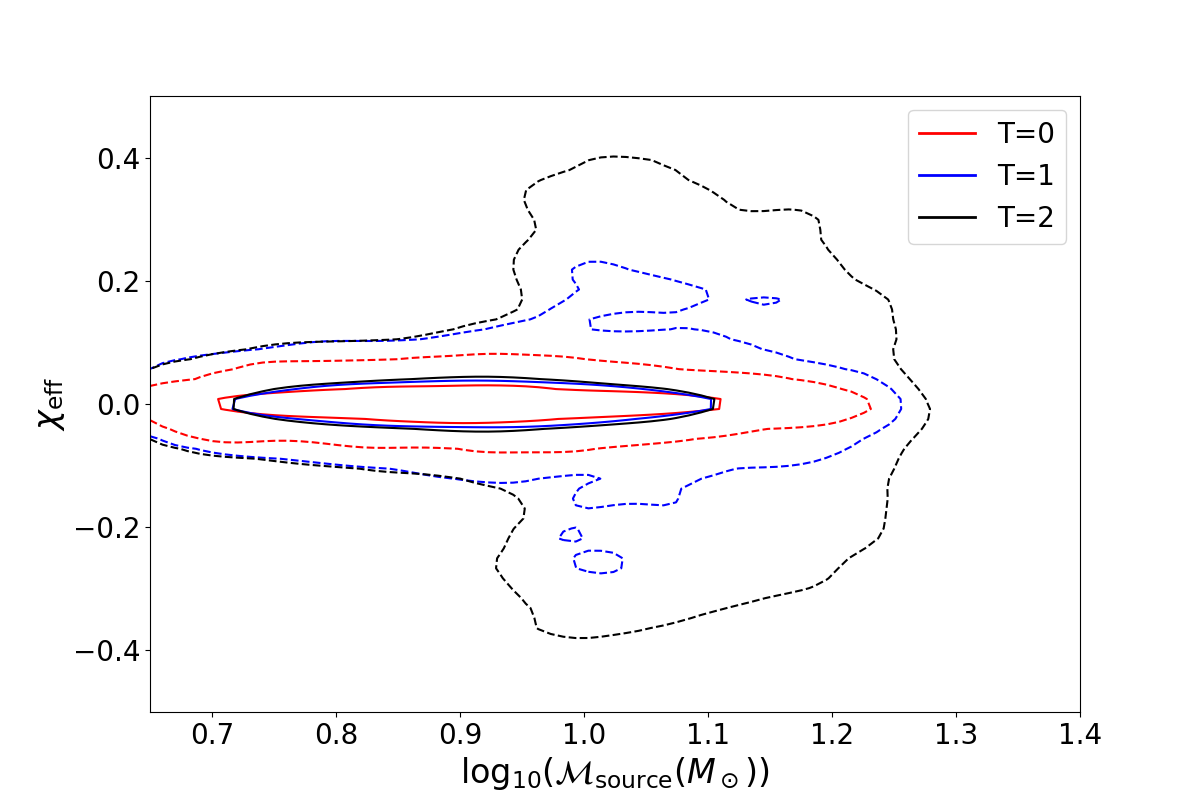}
\caption{\label{fig:timesteps_spin} Joint mass-spin distributions for three successive
time steps.  {\it Top}: The spin 
amplitudes of the more massive merger component versus their masses. {\it Bottom}: The
effective spin parameter $\chi_{\rm eff}$ versus the binary chirp mass.  The contours represent
90\% and 99.9\% contour intervals for mergers at time steps $T=0$ (red), $T=1$ (blue), and $T=2$ (black).}

\end{figure}

As hierarchical mergers occur, a secondary population of high mass, high spin black holes
begins to form alongside the natal population.  In our Monte Carlo procedure, this time 
evolution of the population is reduced to individual time steps, as described in \S\ref{subsec:MC}.
Figure \ref{fig:timesteps} illustrates how the population changes with each time step.  Starting with
a Salpeter IMF with $m_{\rm max}=20M_\odot$ and Beta-distribution spin magnitudes as the natal population (which we
take as our fiducial natal population) we evolve the population
forward for three iterations, allowing 5\% of the black holes to merge at each step and setting the 
coupling strength to $a=2$ and $b=0$.  The red, blue, and black lines show the distributions of the total masses of mergers 
for time steps 0, 1, and 2, respectively.
  
Since the remnant black holes inherit angular momentum from their parents and from their orbit, hierarchical mergers
also produce strong evolution of BH spins \citep{2017ApJ...840L..24F,GerosaBerti}.  
With successive mergers, the total mass distribution tends towards higher masses, and an 
island of high-mass, high-spin black holes begins to grow.  Figure \ref{fig:timesteps_spin} shows 90\% and 99.9\% confidence
intervals for the joint-mass spin distributions at $T=[0,1,2]$. Notably, hierarchical mergers of comparable-mass
binaries introduce a characteristic peak near $\chi\simeq
0.7$, which is why the top panel of Figure \ref{fig:timesteps_spin} shows a surplus of black holes near that spin magnitude.  
Generically, hierarchical mergers
should produce a similar subpopulation of high-mass, high-spin black holes, since general relativity predicts that
a post-merger remnant black hole is always more massive than either of 
its pre-merger components and its final spin is away from zero. The
$\chi_{\rm eff}$ versus chirp mass distribution in the lower panel shows that while $\chi_1$ tends to be large for the hierarchically produced
mergers, the $\chi_{\rm eff}$ distribution is smoothed out around 0, since the black hole spin directions are isotropically distributed.

\subsubsection{Coupling Strength}

\begin{figure}
\includegraphics[width=\columnwidth]{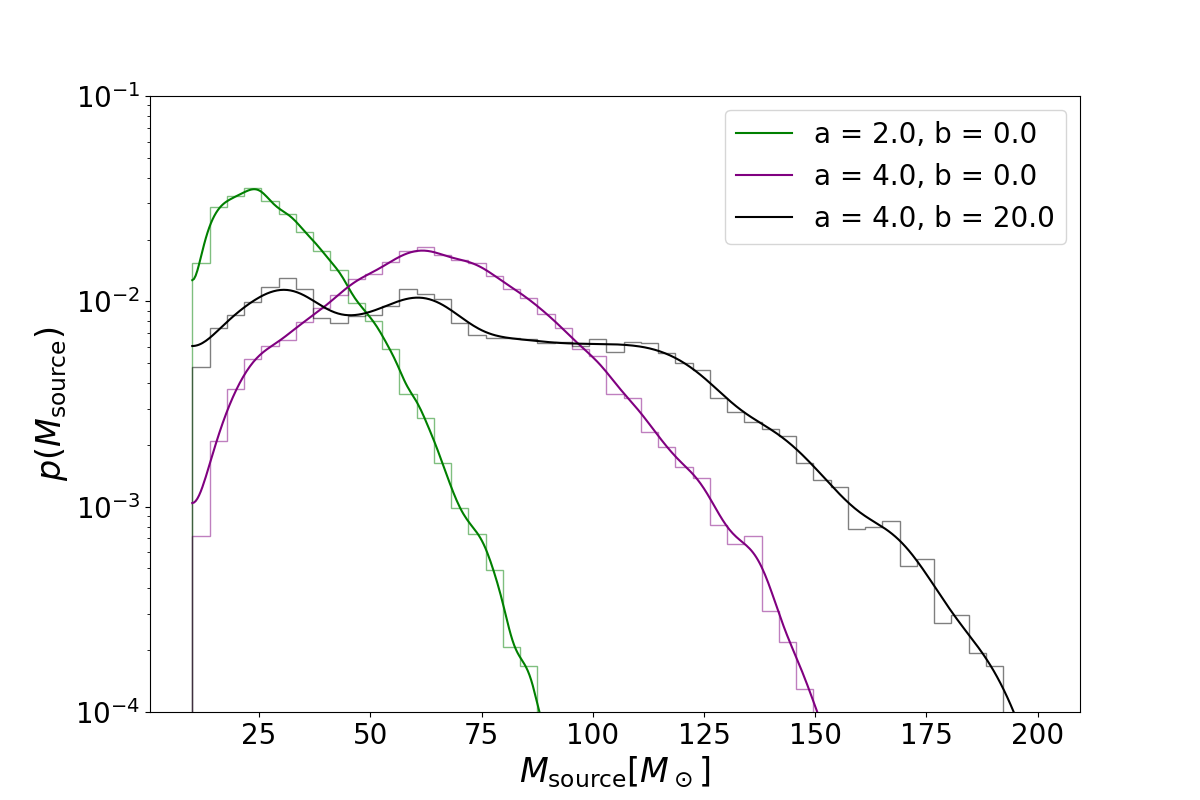}
\includegraphics[width=\columnwidth]{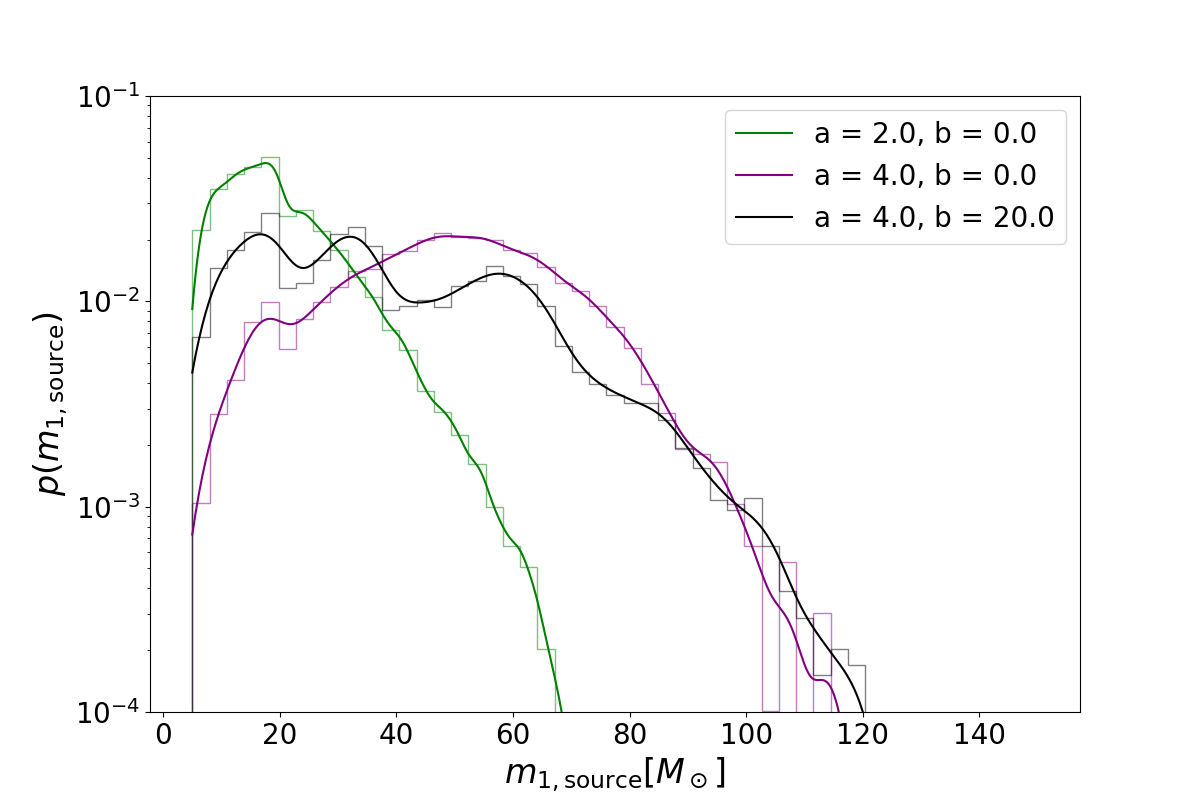}
\caption{\label{fig:masscouplings} Mass distributions for different coupling parameters after four time steps. {\it Top}: The total mass
distribution of mergers for $a=2, b=0$ (green), $a=4, b=0$ (purple), and $a=4, b=20$ (black).  {\it Bottom}:
The distribution of masses of the more massive merger components. The distributions are evolved from the
fiducial Salpeter distribution.}
\end{figure}

\begin{figure}
\includegraphics[width=\columnwidth]{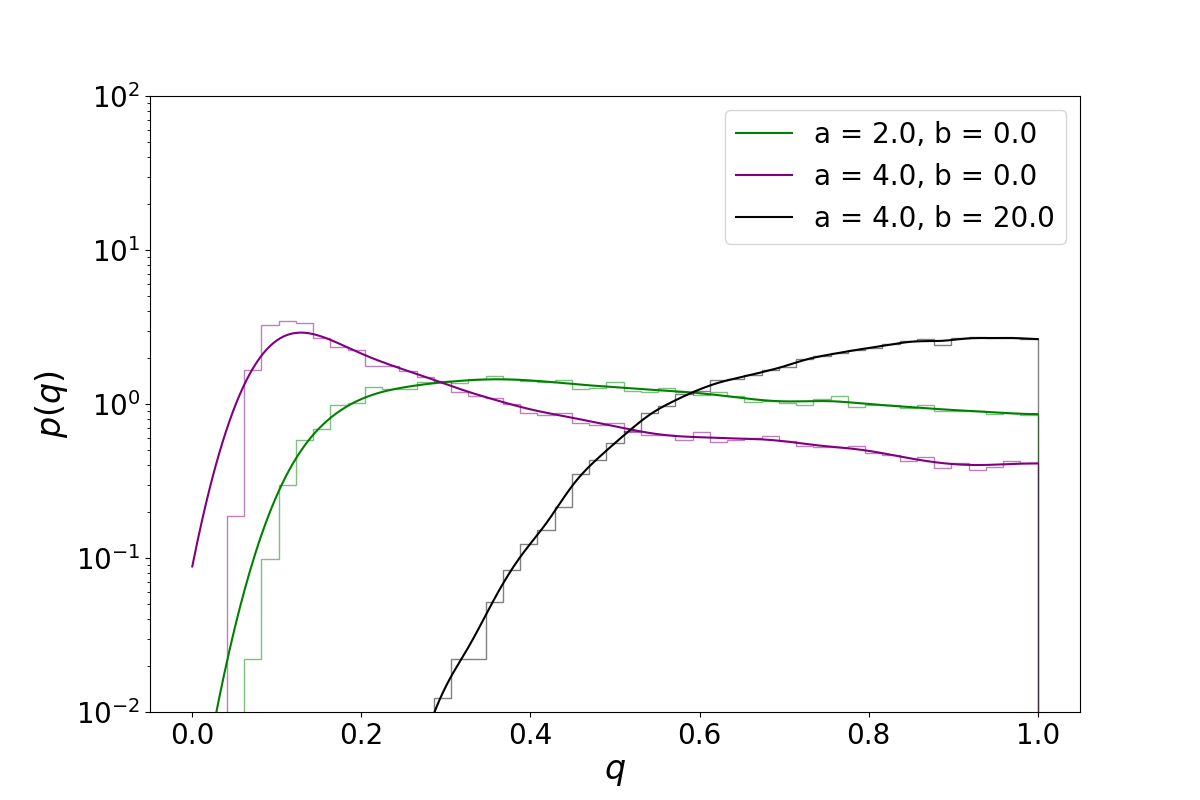}
\includegraphics[width=\columnwidth]{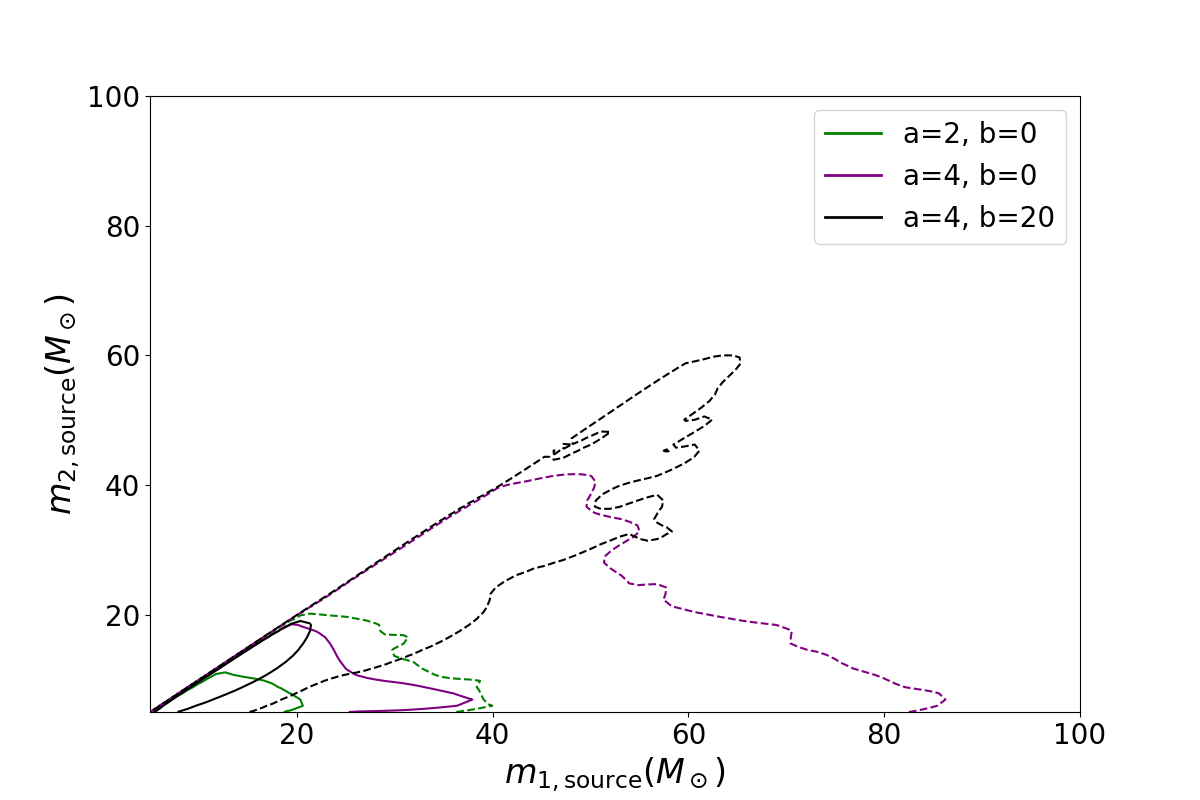}
\caption{\label{fig:couplingsq} The distribution of mass ratios $q=m_2 / m_1$ (top) and component
masses (bottom) after four time steps 
for $a=2, b=0$ (green), $a=4, b=0$ (purple), and $a=4, b=20$ (black), evolved from our fiducial Salpeter
distribution.}
\end{figure}

The overall mass and spin distributions are sensitive to the average coupling strength of black holes.
 Figure \ref{fig:masscouplings} shows the total mass distribution (top panel) and
the primary mass distribution (bottom panel) after four 
time steps for different values of $a$ and $b$.  
Increasing the total mass coupling parameter $a$ drives the most massive mergers to occur, causing the
total mass distribution to quickly expand to higher masses, while increasing the symmetric mass-ratio coupling $b$
simply forces most mergers to be of equal mass components.  Cranking up $a$ and $b$ simultaneously 
gives particularly interesting behavior.  In those cases, the heaviest black holes take place in
mergers, and the products of those mergers are likely to merge again, which can create multiple
distinct peaks in the mass distributions.   As a result, in
the Salpeter natal distribution example, our procedure produces a characteristic ``smoothed staircase'' mass distribution, with
``steps'' in the mass distribution appearing at multiples of the primordial maximum mass $m_{\mathrm{max},0}$.   At very high
mass, these ``step'' features become smoothed out. 

The mass ratio and spin distributions also have characteristic features.  When $a$ is large but $b$ is small, 
a population of highly unequal mass mergers can be produced, as seen in the purple curves of Figure \ref{fig:couplingsq}.
A near-flat mass ratio distribution (shown in green) is found for $a=2$ and $b=0$ in this case, because the natal
mass distribution power law slope ($\alpha=2.35$) is nearly matched to the total mass coupling, so the dearth 
of higher mass black holes is exactly counteracted by their higher likelihood of participating in mergers. As 
$b$ is increased, the distribution begins to favor equal-mass mergers, as shown in the black curve.  

\begin{figure}
\includegraphics[width=\columnwidth]{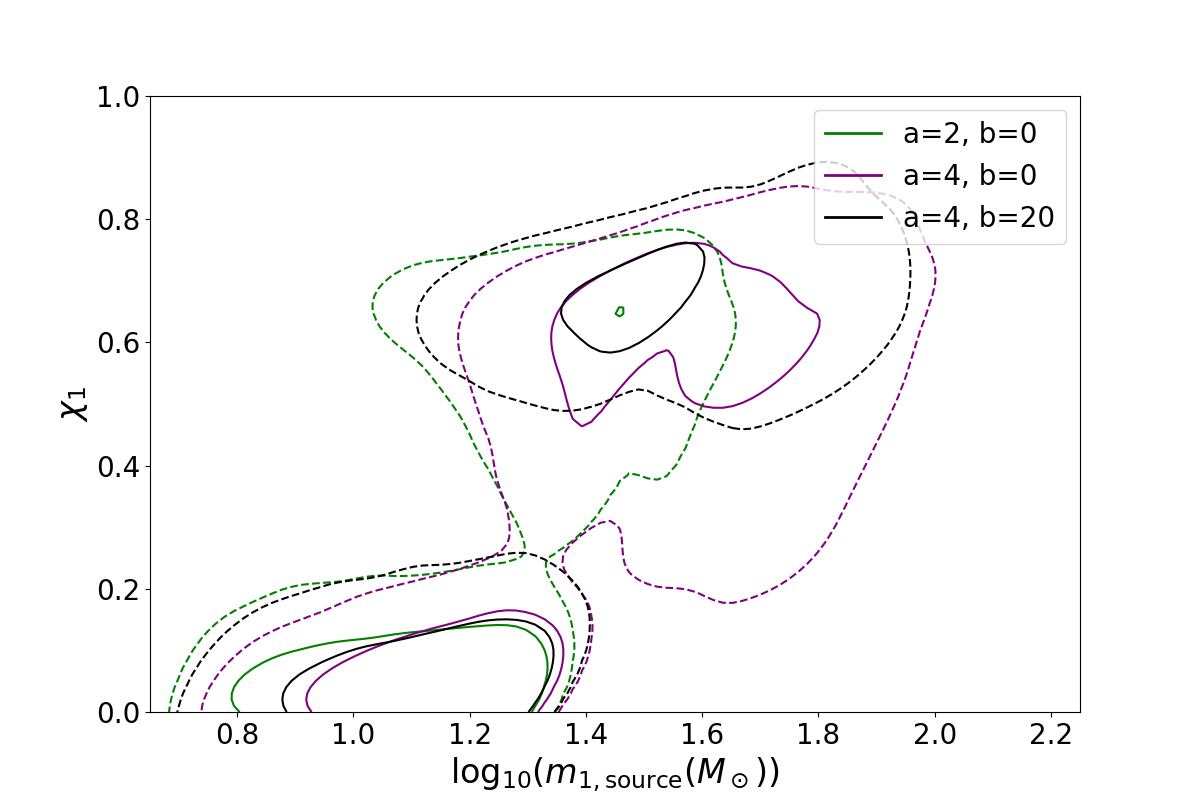}
\caption{\label{fig:couplingsm1chieff} Contours of the joint $m_1$--$\chi_1$ distribution after four time steps 
for $a=2, b=0$ (green), $a=4, b=0$ (purple), and $a=4, b=20$ (black), evolved from the fiducial Salpeter distribution.}
\end{figure}

Figure \ref{fig:couplingsm1chieff} shows contours of the joint primary mass and $\chi_1$ distribution for 
different coupling strengths after four timesteps.  While a high-mass, high-spin subpopulation is
present in all the cases considered here, they are notably affected by the coupling strength parameters.
When $b$ is large, the subpopulation is more concentrated at $\chi_1 \sim0.7$, because the mergers tend
to be equal mass and therefore have a similar final remnant spin.  

\subsubsection{Depletion}

\begin{figure}
\includegraphics[width=\columnwidth]{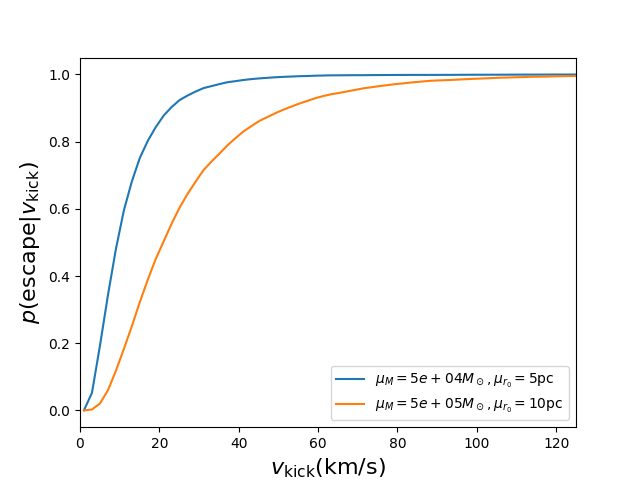}
\includegraphics[width=\columnwidth]{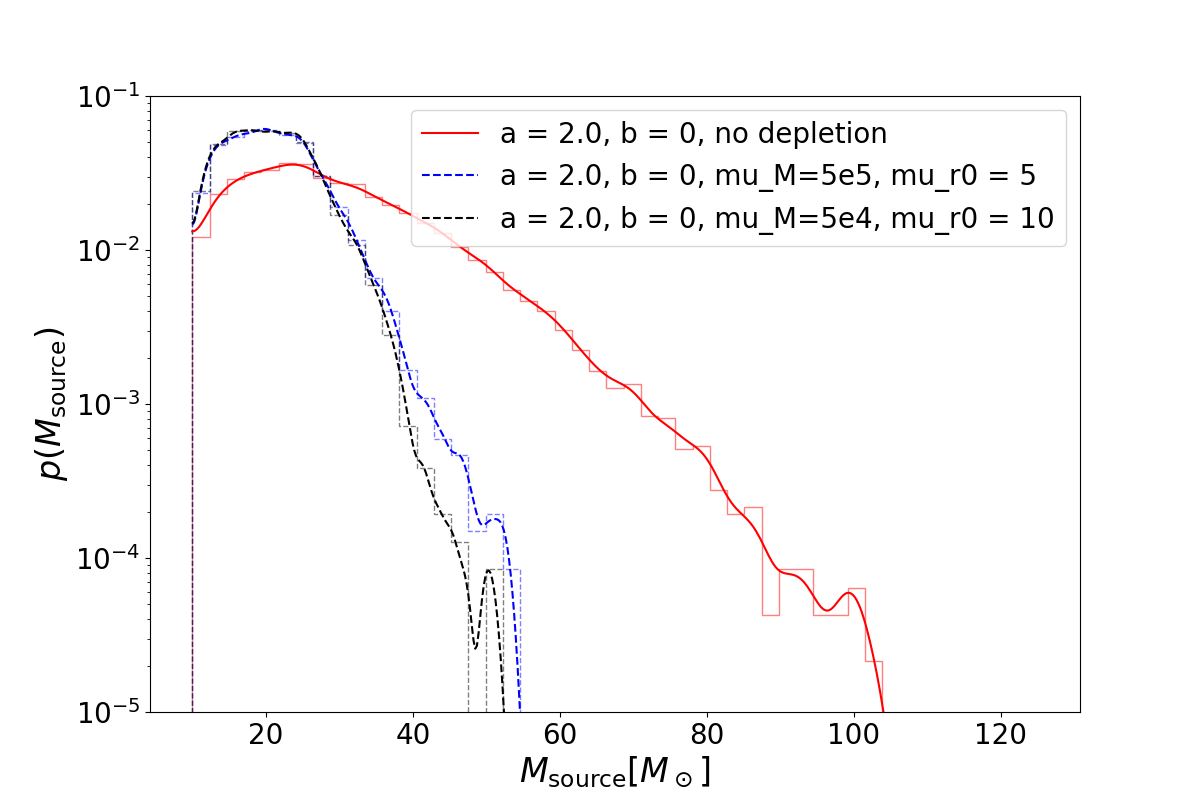}
\caption{\label{fig:depletionvsnodepletion} Escape probabilities and the total mass distribution of mergers for different depletion prescriptions.  
{\it Top}: The escape probability as a function of the kick velocity for ``light'' (blue curve, $\mu_M = 5\times10^4 M_\odot$, $\mu_{r_0} = 10$ pc,  $\sigma_{M}=\sigma_{r_0}=1$)
 and ``heavy'' clusters (orange curve, $\mu_M = 5\times10^5 M_\odot$, $\mu_{r_0} = 5$ pc, $\sigma_{M}=\sigma_{r_0}=1$). 
 {\it Bottom}: The total mass distribution for no depletion (red), ``light'' cluster depletion (black), and ``heavy'' cluster depletion (blue). The
 coupling parameters are set to $a=2,b=0$.}
\end{figure}

The most widely-proposed hierarchical scenario involves hierarchical formation in globular clusters.  Merging
black holes will be very frequently ejected from these low-binding energy environments, strongly suppressing the
prospects for hierarchical mergers through multiple generations \citep{2019PhRvD.100d3027R,2019PhRvD.100d1301G,Favata,Merritt}.  
To illustrate how depletion impacts the observed merger distributions, we incorporate the cluster depletion model from \S\ref{subsubsec:DepletionModel}
into a hierarchical merger population.  Figure \ref{fig:depletionvsnodepletion} plots three total mass distributions, one without
depletion effects, one with ``light'' clusters ($\mu_M = 5\times10^4 M_\odot$, $\mu_{r_0} = 10$pc,  $\sigma_{M}=\sigma_{r_0}=1$), and one with ``heavy'' clusters
($\mu_M = 5\times10^5 M_\odot$, $\mu_{r_0} = 5$ pc, $\sigma_{M}=\sigma_{r_0}=1$).  These hierarchical distributions are
evolved forward four time steps from a fiducial natal distribution under these three depletion prescriptions and with $a=2, b=0$.  
This figure shows that as the confining potentials become shallower, remnant black holes are kicked from the 
environment so that hierarchical mergers are strongly suppressed,
as known from previous work.

\subsubsection{Natal Distributions}

\begin{figure}
\includegraphics[width=\columnwidth]{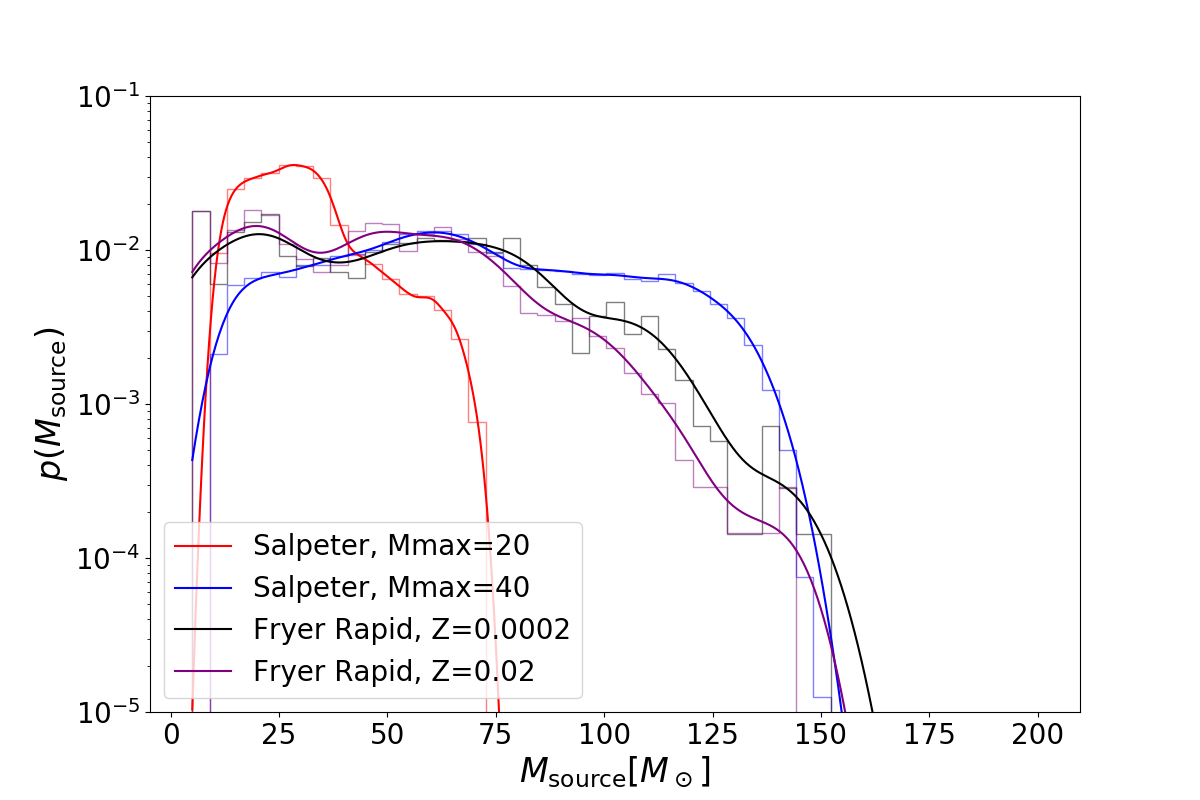}
\caption{\label{fig:diffnatalpops} Total mass distributions after three time steps
for different natal mass distributions.  The coupling strength parameters are
$a=4,b=20$.}
\end{figure}

As we have seen in the previous examples, the hierarchical distributions produced
in our framework contain imprints of the natal populations.  Figure \ref{fig:diffnatalpops} plots
three hierarchical merger total-mass distributions after three time steps assuming the strong coupling
parameters $a=4,b=20$.  Unsurprisingly, the natal distributions with support at higher masses 
quickly evolve to have high-mass mergers.  Additionally, the more complex structure in the
Fryer natal mass distributions is imprinted in the evolved hierarchical distributions, while the
Salpeter-based mass distributions are more smoothed out.  In sum, the natal distribution is
crucially important to the evolution of the mass distribution when hierarchical mergers can 
take place.

%% file: inference.tex

We use an updated version of the \textsc{PopModels} population inference code \citep{gwastro-PopulationReconstruct-Parametric-Wysocki2018} to compare our hierarchical formation model
to real GW observations from GWTC-1 \citep{GWTC1}.
For each collection of observations ${\cal D}$, this code evaluates the inhomogeneous
Poisson likelihood
\begin{equation}
  \mathcal{L}(\rate, \Param) \propto
  e^{-\mu(\rate, \Param)}
  \prod_{n=1}^N
    \int \mathrm{d}\param \, \ell_n(\param) \, \rate \, p(\param\mid\Param),
  \label{eq:inhomog-poisson-likelihood}
\end{equation}
where $\ell_n(\param)=p(d_n|\param)$ is the likelihood of data $d_n$ given binary parameters $\param$, $\mu(\rate,\Param)$ is the expected number of detections, $\rate$ is the merger rate, and $\Lambda$ refers to any relevant model parameters:  all parameters needed to characterize our
hierarchical evolution equations, along with the choice of metallicity and initial conditions.
Unlike \citet{gwastro-PopulationReconstruct-Parametric-Wysocki2018}, we evaluate the integrals $\int d\param \ell_n(\param) p(\param|\Param)$ by using Monte
Carlo integration via samples drawn from our hierarchical model $p(\param|\Param)$, combined with an analytic likelihood $\ell_n(\param)$.
%

\begin{deluxetable}{| r | p{0.35\textwidth} |}
\label{table:models}
\tablehead{\colhead{Model} & \colhead{Description}}
\tablecaption{Hierarchical Merger Models Fit to O1/O2 Data}
\tablenum{1}

\startdata
\hline
Model 1 & 
Natal population: power-law in component mass
  \begin{itemize}
    \item $m_{\rm min}=5M_\odot$
    \item $\alpha \in [-3,5]$, uniform
    \item $m_{\rm max} \in [15,50]$, uniform
    \item $\mathbb{E}[\chi] = 0.047$
    \item ${\rm Var}[\chi] = 0.002$
  \end{itemize}
Coagulation parameters:
  \begin{itemize}
    \item $a\in[1,6]$, uniform
    \item $b\in[1,100]$, log uniform
    \item $T\in[0,9]$, uniform
    \item $w=0.05$
  \end{itemize}  \\
\hline
Model 2 & 
Same as Model 1, except $T\in[0,5]$ and
\begin{itemize}
  \item Beta distribution natal spins
  \begin{itemize}
    \item $\mathbb{E}[\chi] \in [0,1]$, uniform
    \item ${\rm Var}[\chi] \in [0.25]$, uniform
  \end{itemize}
  \item Depletion
  \begin{itemize}
    \item $\mu_M = 5\times10^5M_\odot$
    \item $\mu_{r_0} = 5$ pc
    \item $\sigma_M = \sigma_{r_0} = 1$
  \end{itemize}
\end{itemize} \\
\hline
Model 3 & 
Same as Model 2 except
\begin{itemize}
    \item $\mu_M \in [10^5,10^{10}] M_\odot$, uniform
    \item $\mu_{r_0} \in [5,5\time10^5]$ pc, uniform
\end{itemize} \\
\hline
Model 4 & Mixture of Model 2 and Model A of \citet{O2pops}\\
\hline
Model 5 & 
Same as Model 1 except
\begin{itemize}
  \item Fryer rapid SN natal population
  \item Mixture with Model A of \citet{O2pops}
  \item $T=2$
\end{itemize}
\enddata
\end{deluxetable}

\begin{figure*}[htbp]
  \centering
  \includegraphics[width=0.6\textwidth]{%
    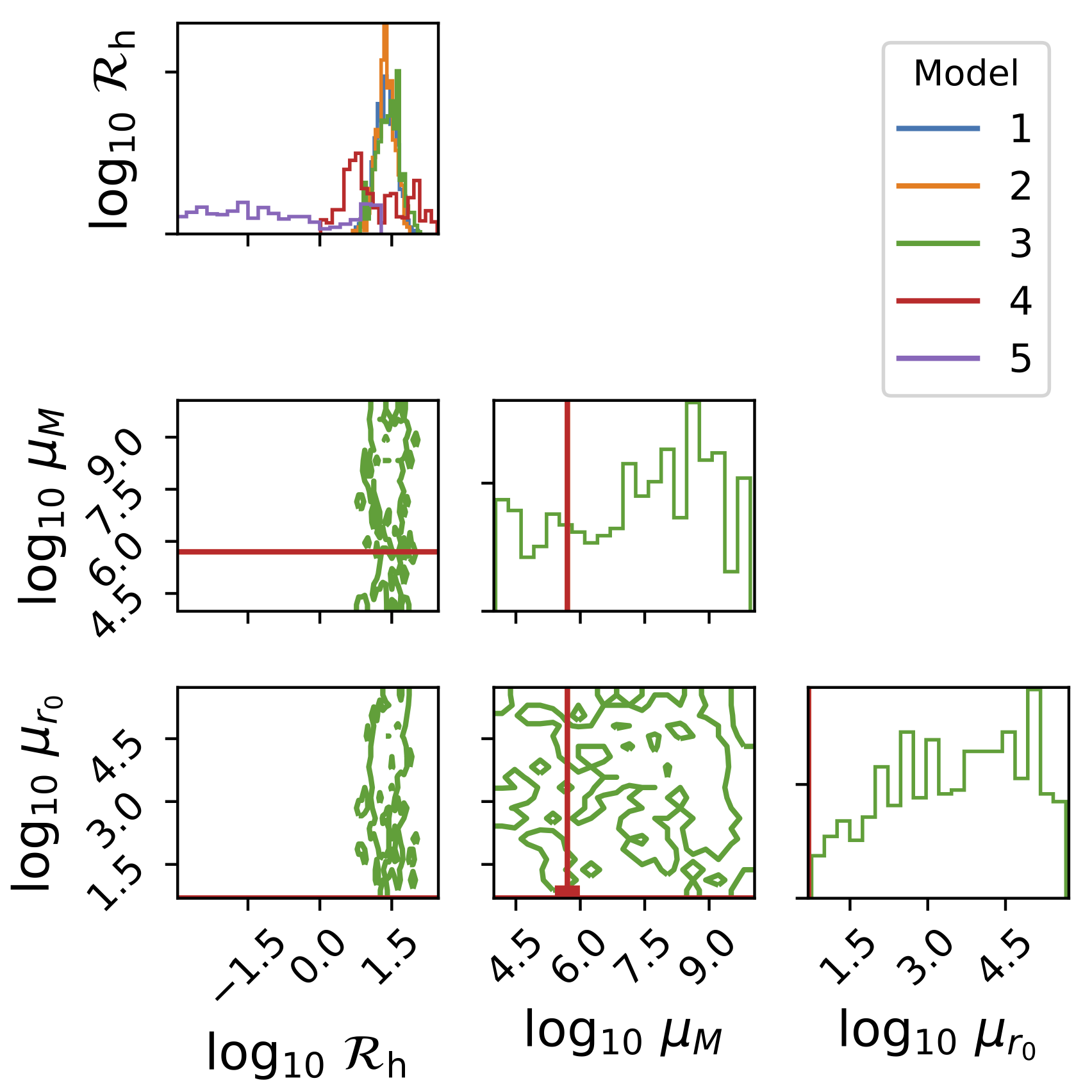%
  }

  \includegraphics[width=0.6\textwidth]{%
    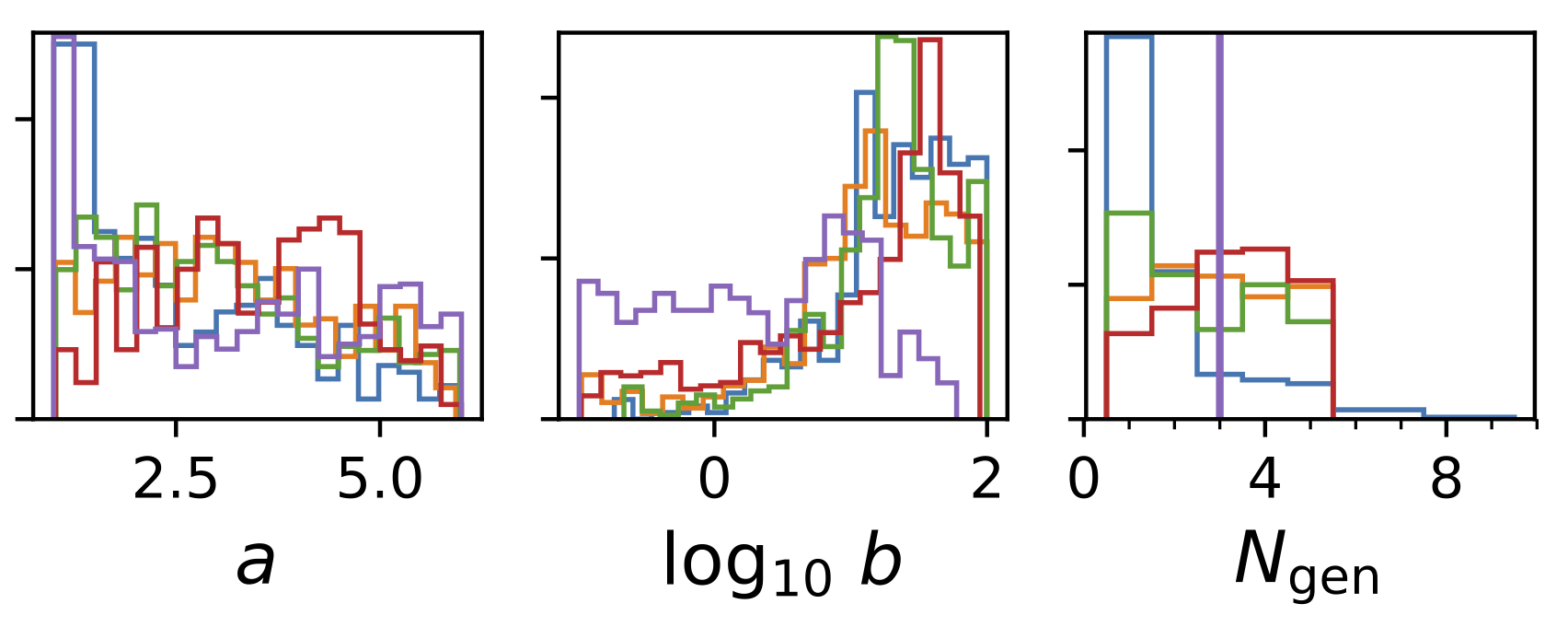%
  }
  \caption{
    Inferred hierarchical parameters with depletion effects, for the 5 models
    listed in Table \ref{table:models}.
    \emph{Top:} Hierarchical merger rates and cluster parameters for
    hierarchical mergers with depletion effects considered.  Note that the
    fiducial model, based on globular clusters vastly underestimates the
    inferred cluster scales.
    \emph{Bottom:} Merger cross section indices for $\Gamma \propto M^a \,
    \eta^b$, for different models both with and without depletion.  Note that
    the only significant difference comes from using the Fryer natal population.
  }
  \label{fig:results1}
\end{figure*}

\begin{figure*}[htbp]
  \centering
  \includegraphics[width=\textwidth]{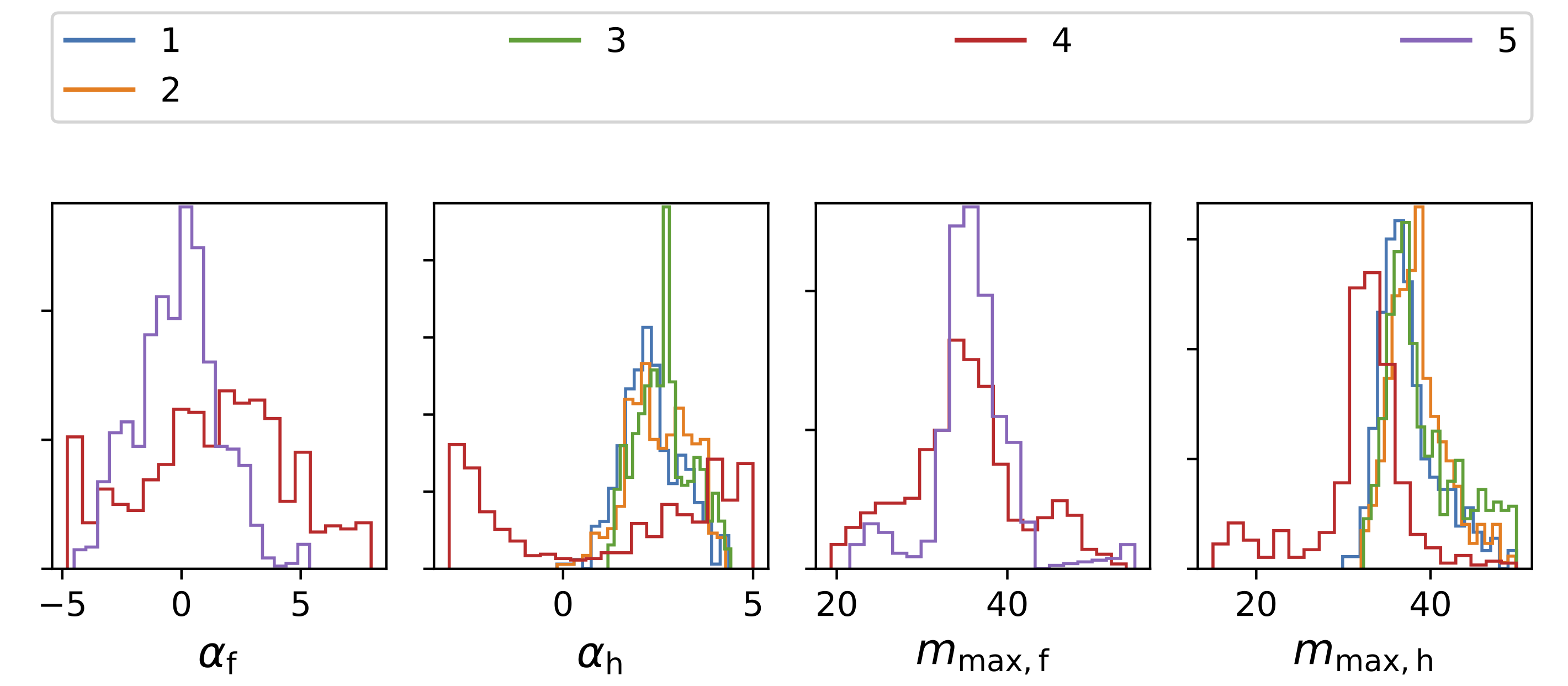}
  \caption{
    Impact of depletion on powerlaw parameters.  Field and hierarchical components
    are denoted with $\mathrm{f}$ and $\mathrm{h}$ subscripts, respectively.
  }
  \label{fig:results2}
\end{figure*}

\begin{figure}[htbp]
  \centering
  \includegraphics[width=\columnwidth]{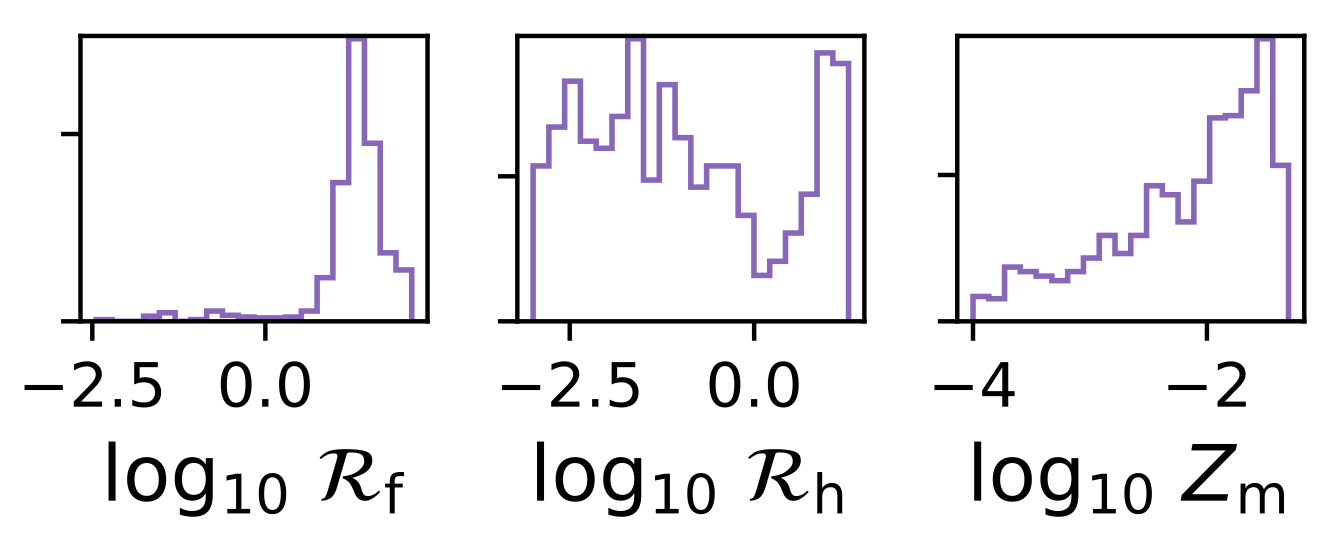}
  \caption{
    Inferred rates and metallicities for model 5. We infer merger rates for both the
  hierarchical component ${\cal R}_{\mathrm{h}}$ and the non-hierarchical component ${\cal R}_{\mathrm{f}}$.
  }
  \label{fig:results3}
\end{figure}

We perform inference with four models, which are described in Table~\ref{table:models}.  The right
hand side describes prescriptions for fixed values and priors in each model.  The posterior
distributions on our inference parameters are shown in Figures \ref{fig:results1}, \ref{fig:results2}, and \ref{fig:results3}.
Model 1 is our most basic phenomenological model, with a power-law-in-component-mass, zero spin natal distribution.  The 
number of iterations and mass coupling parameters are inferred from the data. The blue curves in the right panel of Figure \ref{fig:results1} show
that the data have a slight preference for $a\sim2$ and a strong preference for large $b$ values around $b\sim30$.  The overall rate density of mergers 
in our hierarchical model $\mathcal{R}_{\mathrm{h}}$ shown in the left panel, is consistent with the rates inferred in \citet{O2pops}. Additionally, the natal 
distribution power law index and maximum mass are constrained to similar values to those found in \citet{O2pops}, as 
seen in the blue curves of Figure \ref{fig:results2}.  Given that our hierarchical model reduces to a non-hierarchical model in the 
low-timestep limit and the data favors fewer time steps, it is not surprising that our natal population parameters match the overall population parameters
in \citet{O2pops}.  The inference on the number of time steps is shown in Figure \ref{fig:results1} in terms of the variable $N_{\rm gen}=T+1$, which is the
highest allowed generation of black holes in the population.  

The most widely-proposed hierarchical scenario, however, involves hierarchical formation in globular clusters.  Merging
black holes will be very frequently ejected from these low-binding energy environments, strongly suppressing the
prospects for hierarchical merger \citep{2019PhRvD.100d3027R,2019PhRvD.100d1301G}.  
Model 2 adds a depletion prescription to Model 1 with fixed cluster mass and radius distribution parameters.  In this case,
similar coupling and natal distribution parameters to Model 1 are inferred, which is shown in orange in Figures \ref{fig:results1} and \ref{fig:results2}.  Notably
there is a slight preference for higher total mass couplings $a$ for Model 2 compared to Model 1, because the depletion effects
strongly suppress hierarchical mergers and therefore higher masses from the natal population are favored.  The strong depletion
in this case also results in no preference on the number of time steps.

We then allow the cluster mass and radius distribution parameters to vary in Model 3.  The results of inferring cluster
sizes are shown in Figure \ref{fig:results3}.  Interestingly, the cluster radii and masses are pushed to large
values, far greater than those of real star clusters.  This is partially an artifact of the parameterization chosen here.  
The gravitational potential in the Plummer profile is sensitive only to the ratio of cluster mass to cluster radius, so if we
consider the ratios of $\mu_M$ to $\mu_{r_0}$, the inferred values are roughly similar in gravitational potential to the fixed values used in Model 2.  
In other words, the data prefer somewhat shallow potentials wherein hierarchical mergers are suppressed.
A future parameterization may instead opt for a distribution of gravitational potentials rather than cluster parameters.

Next we consider two mixture models.  In the first (Model 4), we fit a mixture of our Model 2 with Model A from \citet{O2pops}.  Then in Model 5 
we create a mixture of Model A and our hierarchical model applied to the the Fryer rapid SN natal population with no depletion and exactly 3 timesteps of evolution.
\ZD{In these mixture analyses, we simultaneously fit the parameters of Model A (power-law index, maximum mass cutoff, overall rate) and the parameters
of the hierarchical model.}
Figure \ref{fig:results2} shows the distributions of population parameters for the ``field'' (Model A) and ``hierarchical'' mixtures.  The mixtures complicate the picture
significantly. Model 4 (shown in red) in particular has little discerning power on its underlying population parameters due to the additional model freedom.  Model 5 (purple) on the other hand,
for which the natal distribution and number of time steps are fixed, shows some interesting behavior.  In particular, the ``field'' (i.e.~Model A) component parameters are
driven to a near flat distribution in component masses with a slightly lower mass cutoff than for the other models considered inferences.  Meanwhile, the coagulation parameters
$a$ and $b$ are pushed to lower values. These shifts in the inferred parameters are likely due to fixing the number of timesteps to 3 with no depletion.  Fixing the number of timesteps
to 3 favors the existence of some hierarchical mergers which tend to be higher mass.  To counteract the build up of too many high-mass black holes compared to the data, the mass distribution of the field
population is cut off at a lower $m_{\rm max}$ and the coagulation mass coupling is decreased. Also, the inferred metallicity $Z_{\mathrm{m}}$ of the natal population also slightly favors higher values, which 
pushes the natal mass distribution to lower masses, alleviating some of the unwarranted build-up of high mass black holes. Lastly, the contribution of the hierarchical population is subdominant to the field
population, as seen in Figure \ref{fig:results3}.

%% file: discussion.tex
A hierarchical formation scenario provides an efficient way to produce binaries which would otherwise be challenging to
generate: high masses, exceptional mass ratios, and characteristically high spins.  The identification of binaries with
characteristically extreme properties could provide a clear indication of hierarchical formation.    In this section we explore our posterior predictive constraints on these scenarios, within the
framework of the constrained fiducial model described above. We also discuss further extensions of the models presented here and the overall effectiveness of this framework.

\subsection{Posterior Predictive Distributions}

\begin{figure}
\includegraphics[width=\columnwidth]{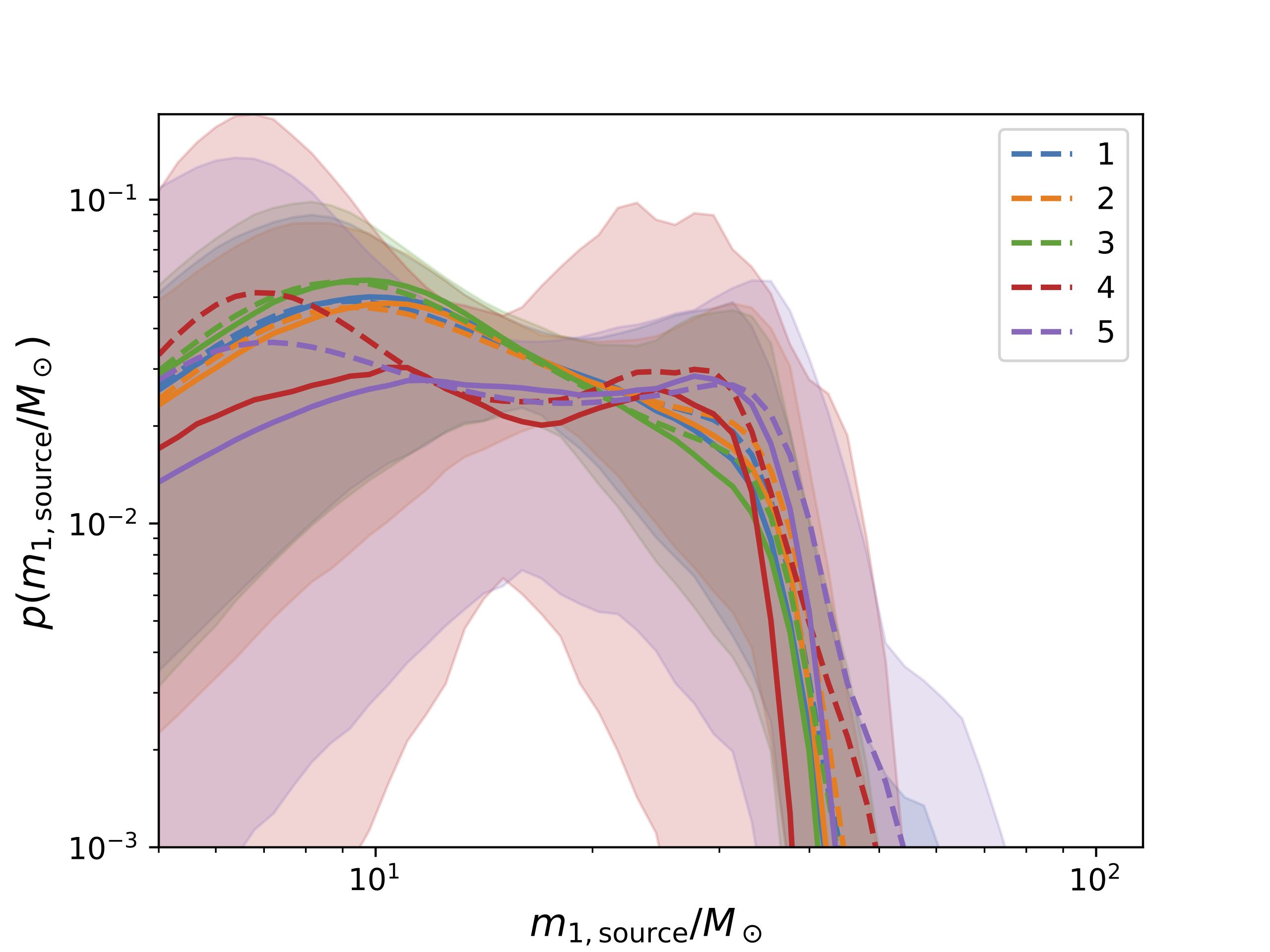}
\caption{
  Inferred $m_{1,\mathrm{source}}$ distributions for Models 1-5.  Shown
  are the median (solid line), posterior predictive (dashed line), and 90\%
  credible intervals (shaded region).
}
\label{fig:ppd:m1}
\end{figure}

Figure \ref{fig:ppd:m1} shows our inferred posterior mass distribution, both intrinsic and detection-weighted, which
resemble the conclusions in \citet{O2pops}.  Specifically, we infer a mass distribution for the more massive component in merging  black holes ($m_1$) that is
approximately a power law between $10 M_\odot$ and $30 M_\odot$, followed by a rapid decrease at higher mass.   Notably,
this figure shows characteristic decay and ``echo'' features at about $30 M_\odot$, inherited by our formation model; these
features could be probed by future observations and used to better constrain
hierarchical formation.

\begin{figure}
\includegraphics[width=\columnwidth]{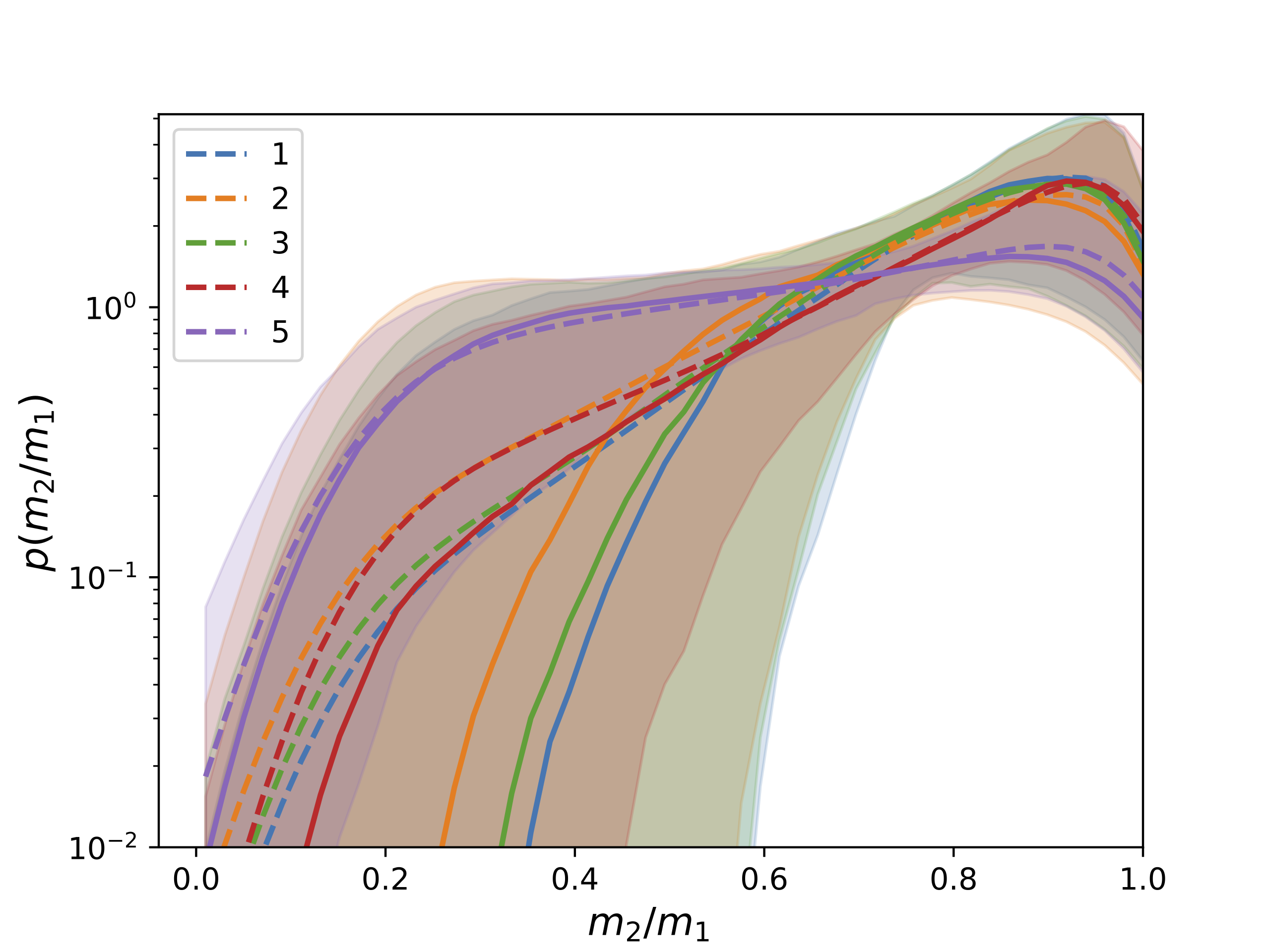}
\caption{
  Inferred $m_2 / m_1$ distributions for Models 1-5.  Shown
  are the median (solid line), posterior predictive (dashed line), and 90\%
  credible intervals (shaded region).
}
\label{fig:ppd:q}
\end{figure}

Figure \ref{fig:ppd:q} shows our inferred mass ratio distribution.  Because GWTC-1 does not include a significant
component of asymmetric binaries, our posterior necessarily strongly favors hierarchical models which preferentially
produce binaries with $q\simeq 1$.   Constraints on binary mass ratios will very strongly constrain prospects for
hierarchical formation, particularly insofar as some hierarchical scenarios produce significant numbers of highly
asymmetric mergers \citep{2019arXiv190609281Y}.


%
\subsection{Possible Extensions}
In this article, we have shown a few possible model choices and prescriptions, but as we have emphasized, many other choices
could be made.  For example, our parameterizations of the coagulation coupling and cluster depletion are essentially 
phenomenological, but future work could incorporate the results of N-body simulations which evolve clusters of stars
and black holes as well as incorporate observational constraints on star clusters.  

Our current parameterization also assumes that there is no evolution of the rate or mass and spin
distributions with redshift.  Given the evolution of the cosmic star formation rate, it is likely that the rate of 
black hole mergers is increasing between $z=0$ to $z\sim 1$, and analysis of available gravitational wave data has 
already lent weak support to that hypothesis \citep{2017ApJ...840L..24F, O2pops}.  Additionally, properties of the environments
in which black holes merge (such as the distribution of cluster potentials) could have changed over cosmic time, leading to observable differences in the mass and spin distributions
between low and high redshifts.

Another possible extension to our model would be to consider more complex mixtures of populations. We briefly considered
a ``field'' plus ``cluster'' mixture population here, but if mergers are occurring in AGN disks, globular clusters, in the field, 
and from a primordial population, more mixture components would need to be added. More gravitational-wave data will be needed
before embarking on such investigations, as the number of parameters of such a complex mixture will proliferate.

Lastly, we note that this work has not considered neutron stars. After this work reached maturity, we became aware of 
a similar investigation targeting hierarchical formation of neutron stars \citep{Gupta2019}. Nevertheless, our framework could neatly incorporate 
neutron stars by substituting in a neutron-star natal distribution and a model for neutron-star merger remnant masses,
spins, and kick velocities.  The main new addition in a hierarchical population based on NS mergers would be the need
to incorporate an equation of state.  

\subsection{Efficacy of this Hierarchical-Merger Population Framework}

Our phenomenologically-parameterized framework provides an efficient way to characterize the contribution of
hierarchical mergers to a compact binary population, and to interpret BH mass measurements as constraints on this
sub-population.  We can use GW measurements to infer the natal mass and spin distribution, as well as evolution parameters.
Of course, our model cannot completely disambiguate these two features without other observational or physical input.
As a trivial example, any set of GW observations can be explained by a non-hierarchical population and a
suitably-overfit natal mass and spin distribution.  
If, however, physical constraints limit the flexibility of the natal
BH binary distribution to populate parts of parameter space, then the presence of merging BHs in those distinctive
regions provides evidence for hierarchical formation.  In such a scenario, our framework enables us to provide first
constraints on a hierarchical merger interpretation.

In this work, motivated by LIGO's observations in GWTC-1, we have emphasized formation scenarios with strong effective
coupling, to produce a binary black hole population which favors comparable-mass mergers.  We expect that more
theoretically-motivated choices for these interaction exponents will favor a wider range of mass ratios.  As noted in
previous work \citep{gwastro-agndisk-McKernanPredictMassSpin2019}, high mass ratio binaries could  be a distinctive
signature of certain hierarchical growth scenarios.   The presence or absence of high mass or high-mass  ratio binaries strongly
constrains our model parameters and the overall hierarchical merger rate.

Another characteristic feature of some hierarchical merger scenarios is a ``smoothed staircase'' or multi-modal pattern in the mass
distributions.  In the simplest case where the natal population is just composed of black holes with mass $M_{\rm natal}$, ``harmonics''
of the natal mass should appear in the black hole mass spectrum at multiples $M_{\rm natal}$.  If the natal distribution is 
sufficiently complex, such harmonics may be blended out, but as shown in \ref{sec:model}, there are intermediate cases
where smoothed out harmonics or ``staircases'' are still noticeable.  

Conversely, many mass and spin distributions cannot be naturally produced from hierarchical evolution.  If
hierarchical formation is proposed to explain a subpopulation of 
high-spin or high-mass or high-mass ratio binaries, then the relative merger rate of this feature is often bounded
above.  For example, we would need sufficient numbers of  low-mass BHs to explain a population of high-mass, high-spin
BHs entirely through hierarchical formation.

Once an observed population is fit with a realization of a hierarchical population, our Monte Carlo method enables
computation of some interesting quantities.  In principle, each Monte Carlo sample has an associated ``family tree'' which tracks
successive mergers that the black hole had previously undergone\footnote{The evolutionary tree of each black hole
is not tracked in our implementation of the Monte Carlo method, but future upgrades to the code will integrate this 
tracking.}.  Thus one can evaluate the probability that a given black hole was hierarchically formed and underwent $n$ previous
mergers.  Alternatively, upper limits can be set on the rate of hierarchical mergers if none are suspected in the 
GW sample.

%% file: conclusion.tex
Observing hierarchically formed black holes is an exciting prospect for gravitational-wave detectors.  We have 
presented here a self-consistent framework for generating black-hole merger populations that includes
hierarchical formation. This framework evolves arbitrary natal black hole populations, enabling any
existing black hole distributions to be extended to include hierarchical mergers. With the cases we explore here, our fits suggest that scenarios with many hierarchical
mergers are disfavored.  

In this work, we simulated coagulation and depletion effects while assuming
 the binary black hole population is not continuously repopulated from another reservoir of black
holes.   We will explore self-consistent repopulation in later work.
We also perform simplified averaging, not allowing for a distribution of initial conditions like metallicity or for
trends versus redshift.  Our scheme ignores higher order correlations and multi-body effects, thus 
averaging everything into an effective cross section which is constant.  
Our approximation is reasonable in the limit of weak hierarchical reprocessing dominated by a low-mass seed population; we
defer more sophisticated averaging to future work.
Additionally, we adopt a simple dependence of  $\Gamma$
on total mass, allowing it to increase without bound according to a single power law  as
the binary mass increases.  More detailed investigations will produce more complex dependence of $\Gamma$ on mass.
The results of our inference on GWTC-1 with partially constrained versions of our model framework show consistency with
\citet{O2pops}, but more detections are required to make more definitive statements about whether hierarchical
formation of black holes is at work. \ZD{Our fits to GWTC-1 hint that hierarchical merger scenarios are 
not required to fit the population, but are also not ruled out by the population.  However, the conclusions
drawn herein are subject to the simplifying assumptions made for this preliminary analysis.} 
Incorporation of more astrophysically motivated inputs to the framework
will be necessary to put the tightest constraints on the formation environments and scenarios.  With the wealth
of black hole merger detections we expect to see in the coming years, the prospects for uncovering hierarchically
formed black holes are promising, and the framework we have presented herein is well-suited for such investigations.\\

The general purpose hierarchical population code will be released for public use in the near future.

%% file: appendix_semianalytic.tex
\begin{figure}
\includegraphics[width=0.45\columnwidth]{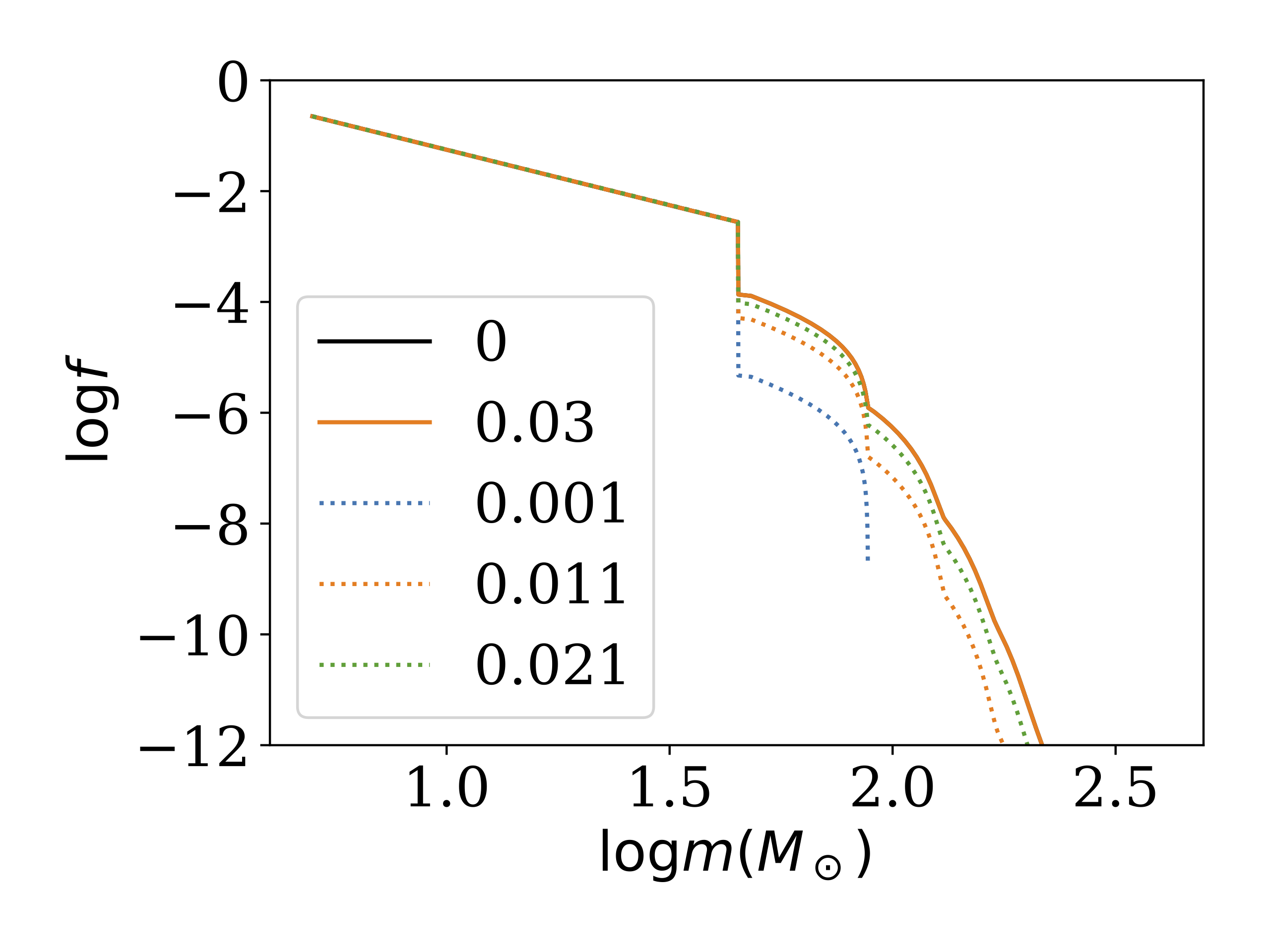}
\includegraphics[width=0.45\columnwidth]{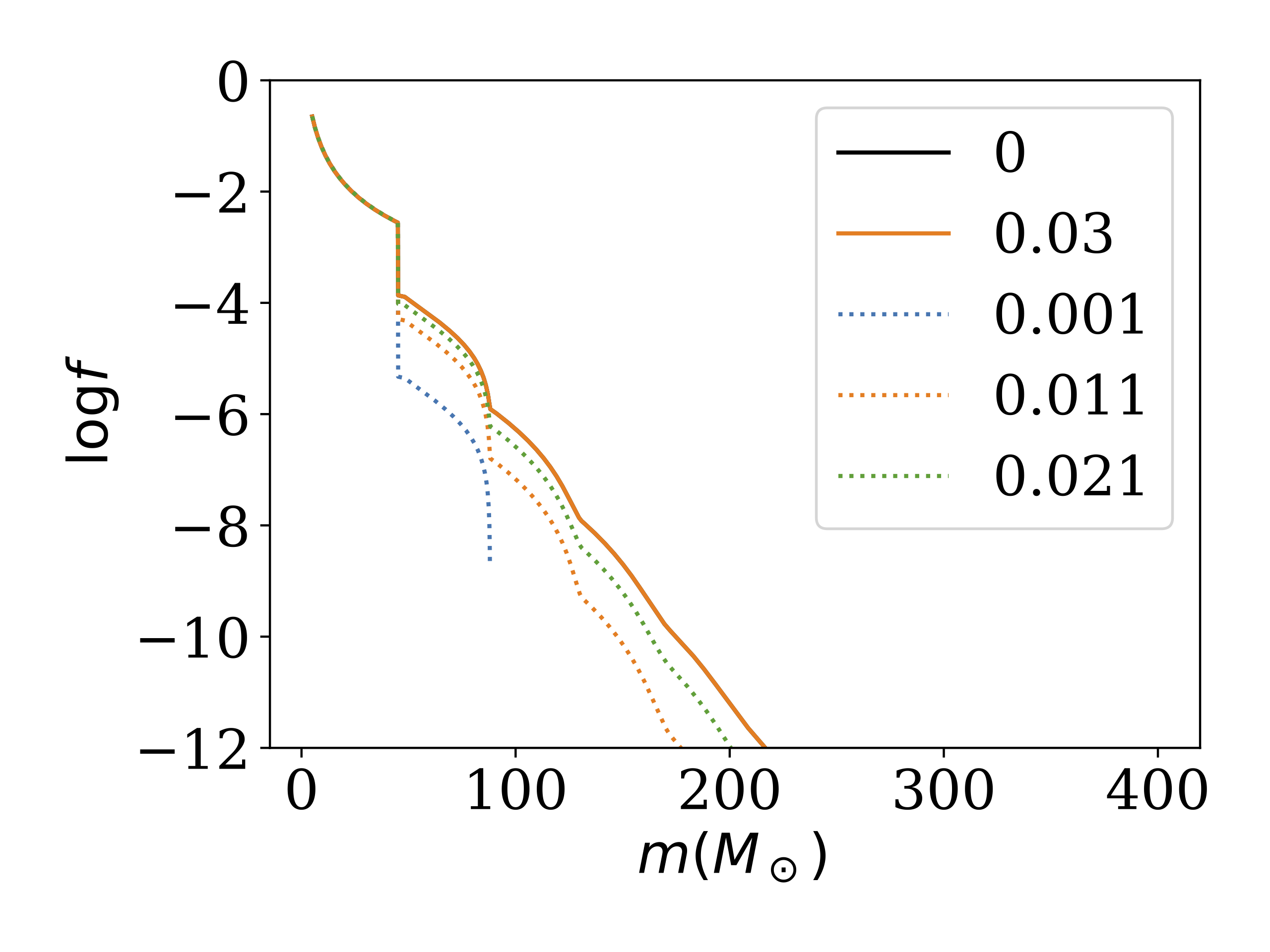}
\caption{\label{fig:IllustrateMass}Synthetic mass distribution as a function of $x$ due to hierarchical mergers with
 $a=2,b=0$ and no depletion,
 starting with a truncated power law distribution at $x=0$.  Colors and legend denote different choices for $x$.  Bottom panel uses a
 log-linear scale to highlight exponential decay at large mass. }
\end{figure}

In the text, we consistently employ a concrete Monte Carlo implementation of hierarchical mergers.  This powerful method
allows us to efficiently incorporate the best merger physics, but uses discrete generations.  In this appendix, for pedagogical purposes we provide a
toy model implementation of a true continuous-time coagulation equation.  In this approach, we consider only the
evolution of binary mass, assuming no mass is lost during mergers; we neglect spins and depletion.   After these
simplifications, our model is essentially analytically tractable, and can be understood by both perturbation theory and
direct numerical simulation.    In this appendix, we present a few supplementary illustrations of these hierarchical
calculations, to further illuminate our model's behavior at very high mass and in the absence of depletion.

As our first example, to illustrate the parameters $\zeta$ and $a$, Figure \ref{fig:IllustrateMass} shows the results of evolving Eqs \ref{eq:coag} starting with an initial power law mass distribution through different
ranges of interaction parameter $\zeta \ll 1$, for two choices of $a$ and for $b=0$.  The dotted curves show the results of a direct numerical
time integration; the solid curves show our Monte Carlo procedure; and different colors indicate different choices for
$x$ and $a$ respectively.

This example first shows how the parameter $\zeta$ controls the effective number of generations at the reference
parameters, absorbing factors present in the overall interaction time $T$ and in the interaction cross section
$\Gamma$.   
As expected based on perturbative arguments, higher-order generations increase in significance in proportion to $x^g$
for $g$ the number of generations.  
At very high mass, the hierarchical mass distribution approaches an exponentially decaying function of  $m$, which  increases exponentially with
$x^2$.
\footnote{Using an ansatz $f(m,x)=g(x)e^{-Am}m^z$, we can see a stationary exponential high-mass  solution exists
 for $b=0$.}

Second,  this example shows how the coagulation equation successively reprocesses each generation, potentially with
different interaction scales.   In this example and generally in the usual case that
$a,b>0$, binaries with smaller masses or more asymmetric mass ratios by construction interact even less frequently.
Conversely, within our framework high mass binaries rapidly undergo multiple generations of mergers.  As a result, in
this power law example, our procedure produces a characteristic ``smoothed staircase'' mass distribution, with
``steps'' in the mass distribution appearing at multiples of the primordial maximum mass $m_{max,0}$.   At very high
mass, these ``step'' features become smoothed out. 

Third, this example shows the importance of the interaction cross section: because we adopt $b=0$ (no preference to any
mass ratio) and because low-mass BHs are dramatically more prevalent than high-mass binaries, the overall merger rate
$f(x)f(x')\Gamma_{x,x'}$ for binaries with one massive component  $x$ is  overwhelmingly dominated mergers where $x'$ is
drawn from this scenario's ubiquitous low-mass black holes.  As
a result of these frequent minor mergers, features in the mass spectrum proportional to the primordial maximum mass are
rapidly smoothed out, both in $x$ and as we go to higher multiples of the maximum mass, except for the first feature.

In sum, our coalgulation model naturally ``builds up'' self-consistent hierarchical populations, producing mass
distributions which can (but need not) possess clear features reflecting the number of generations and any sharp cutoffs
present in the seed distribution.

\begin{figure}
\includegraphics[width=\columnwidth]{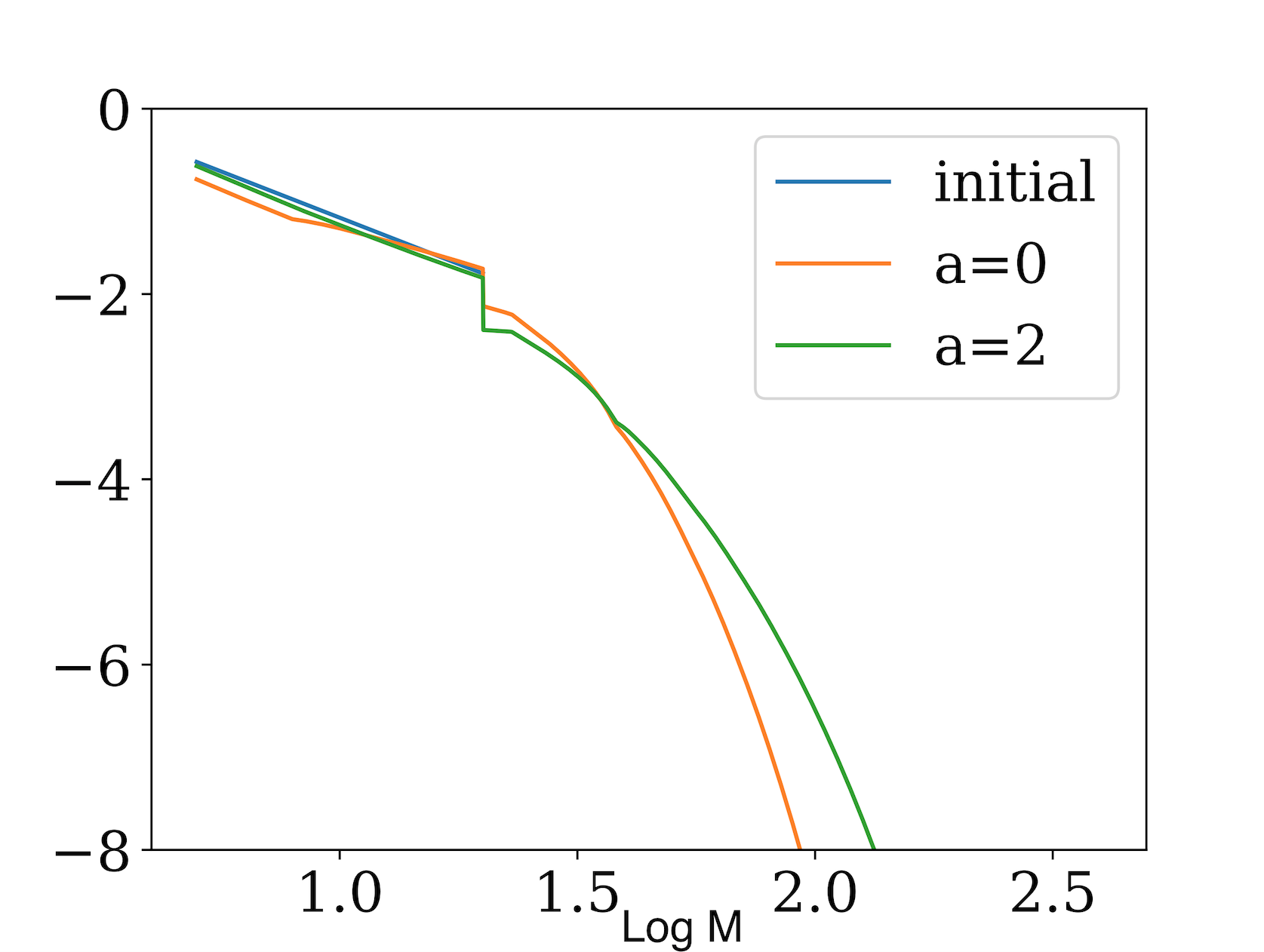}
\caption{\label{fig:IllustrateMass:2}Illustration of the evolving mass distribution: logarithm of the mass spectrum versus mass.}
\end{figure}

As our second example, we consider how features of the high-mass mass distribution are inherited from the low-mass mass
spectrum., using a broad, featureless power law distribution
initially $f(m)\propto m^{-\alpha}$ at low mass.    For simplicity and unlike the example used above, we consider interactions with $b \gg 1$, insuring that
almost all mergers occur between \emph{comparable}-mass binaries.  

Qualitatively speaking, coagulation requires the formation rate of BHs with mass $2 m$ must be $\Gamma f(m)^2\propto
m^{a-2\alpha}$, a slope which can be shallower or steep than the low-mass slope $(-\alpha)$ depending on the sign of
$a-\alpha$.  Evidently, as corroborated by Figure \ref{fig:IllustrateMass:2},  larger $a$ favors higher-mass black holes
and a more extended tail in the mass distribution.  As in the previous example, at very high masses the distribution 
decays exponentially, with a coefficient that depends on $a$. 

%% file: merge_frac.tex
Rather than use a continuous-time coagulation equation, which implicitly allows BH remnants to participate in subsequent
hierarchical mergers immediately, we employ discrete timesteps with a specific fraction 
 $w$ of BHs that participate 
in  mergers at each iteration.  As $w\rightarrow 0$, our algorithm
converges to continuous coagulation, because the population changes 
slowly over many iterations, ensuring that $\Delta f(x,t)/\Delta t$ is small at
each step and that post-merger remnant black holes are immediately available to 
merge again.  As $w$ increases, our iterative process increasingly differs from continuous evolution. 
  In effect, $w$ encodes a ``recycling delay time," i.e.~the time for 
a post-merger remnant black hole to be re-integrated into the population.  We emphasize that while
larger $w$ loses fidelity to the continuous coagulation equation, high $w$ can still model a real
compact object population. For example, it is conceivable that all natal black holes merged 
at an early time (i.e.~$w$=1), and then all participate in second generation mergers
at later times.  

\begin{figure}
\includegraphics[width=\columnwidth]{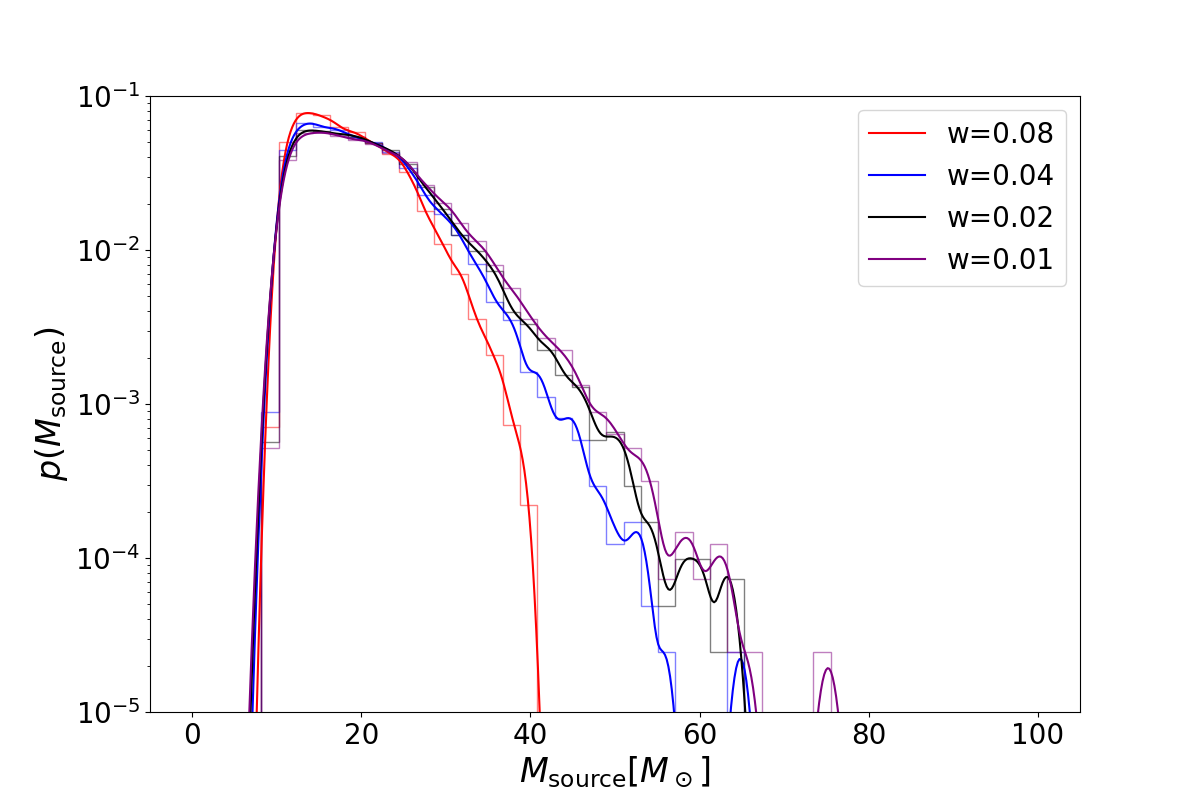}
\caption{\label{fig:mergefrac_fixedtime} The total mass distribution for different fractions
of mergers per time step.  The number of time steps $T$ is varied such that the total number
of mergers is constant.  Specifically, $T=2,4,8,16$ for $w=0.08,0.04,0.02,0.01$, respectively.}
\end{figure}

As a concrete example, we apply our algorithm to our fiducial power-law natal population using different $w$ values
while holding constant the total number of mergers.  In other words, we ensure $wT$ is a constant, 
where $T$ is the number of iterations of our Monte Carlo method.  The coupling constants $a$ and $b$ are set to 0.  
Figure \ref{fig:mergefrac_fixedtime} shows the
total mass distribution of mergers after
$T=2,4,8,16$ iterations with $w=0.08, 0.04, 0.02, 0.01$, respectively.  At lower masses, the distributions
roughly agree, but only the small $w$ cases have tails extending to higher masses, since those cases are 
closer to continuous coagulation wherein there is no recycling delay time and post-merger remnants are 
free to re-merge immediately.  

